\documentclass[final,5p,times,twocolumn]{elsarticle}

\usepackage{graphicx}
\usepackage{amsmath}
\usepackage{amssymb}
\usepackage{array}
\usepackage{booktabs}
\usepackage{algorithm}
\usepackage{algorithmic}
\usepackage{listings}
\usepackage{color}
\usepackage{url}
\usepackage{xcolor}
\usepackage{hyperref}
\usepackage{multirow}
\usepackage{longtable}
\journal{Ocean Engineering}

\begin{document}

\begin{frontmatter}



\title{A Comprehensive Review of Phase-Averaged and Phase-Resolving Wave Models for Coastal Modeling Applications}

% --- Authors ---

\author[inst1]{Md Meftahul Ferdaus}
\ead{mferdaus@uno.edu} % Email for corresponding author

\author[inst1]{Nathan Alton Cooper}
\author[inst1]{Austin B. Schmidt}
\author[inst1]{Pujan Pokhrel}

\author[inst2]{Elias Ioup}

\author[inst1]{Mahdi Abdelguerfi}

\author[inst2]{Julian Simeonov} 

% --- Affiliations ---

\affiliation[inst1]{organization={Canizaro Livingston Gulf States Center for Environmental Informatics, Department of Computer Science},
            addressline={The University of New Orleans}, 
            city={New Orleans},
            postcode={70148}, 
            state={Louisiana},
            country={USA}}

\affiliation[inst2]{organization={Center for Geospatial Sciences, Naval Research Laboratory},
            addressline={Stennis Space Center}, 
            city={Hancock County},
            state={Mississippi},
            country={USA}}

%% Abstract
\begin{abstract}
Predicting ocean wave behavior is challenging due to the difficulty in choosing suitable numerical models among many with varying capabilities. This review examines the development and performance of numerical wave models in coastal engineering and oceanography, focusing on the difference between phase-averaged spectral models and phase-resolving models. We evaluate the formulation, governing equations, and methods of widely used third-generation phase-averaged spectral models (SWAN, WAVEWATCH III, MIKE 21 SW, TOMAWAC, and WAM) alongside advanced phase-resolving models (FUNWAVE, SWASH, COULWAVE, and NHWAVE) that employ Boussinesq-type equations and non-hydrostatic formulations. The review begins with early parameterized models and progresses to contemporary third-generation models, which solve the wave action conservation equation with few spectral constraints. A comparison of the models’ efficiency, accuracy in nearshore conditions, ability to resolve nonlinear wave-wave interaction, simulate wave breaking, diffraction, and wave-current interactions is provided. Applications in operational forecasting, extreme event simulation, coastal structure design, and assessing climate change impacts are discussed. The validation of these models and the statistical metrics and intercomparison studies used are addressed. A discussion of the limitations in computational scalability, physics parameterization, and model coupling is provided, along with emerging trends in high-resolution modeling and hybrid models. This review guides researchers in evaluating which model(s) to use in coastal and oceanographic research.
\end{abstract}


% \begin{highlights}
% \end{highlights}

%% Keywords

\end{frontmatter}



\section{Introduction}
Ocean waves influence coastal and oceanic processes, affecting sediment transport, shoreline development, coastal flooding, offshore structure loadings, and maritime operations \cite{Cavaleri2007}. More than 40\% of the world's population lives near coastlines, and the economic value of coastal-related resources is trillions. Therefore, accurate prediction of wave conditions is essential for managing coastal zones, designing infrastructure, ensuring safe navigation, and planning for climate adaptation \cite{Lavidas2018,Karjoun2023}. Recent hurricanes, typhoons, and storm surge damage demonstrate the need for strong numerical wave models to help forecasters, assess hazards, and project future climate changes \cite{Irham2024,Yao2025}.

In the past fifty years, numerical wave modeling has advanced from basic models to sophisticated computational systems that simulate complex wave phenomena across varied spatial and temporal scales \cite{WAMDIG1988,Roland2014,CasasPrat2024}. These advancements are due to progress in wave physics, computational power, observation methods like satellite altimetry, and numerical algorithms \cite{MaitlandDavies2022,Alday2022}. Modern wave models can accurately resolve complex processes such as non-linear interactions, wave-current coupling, depth-induced breaking, diffraction, and energy dissipation \cite{Delpech2024,Fujiwara2024,Thomas2023}.

Wave numerical models are categorized as phase average (or spectral wave) models and phase resolution models based on how they treat the wave phase \cite{Thomas2015, Reidulff2023}. Spectral wave models represent the time evolution of the wave energy spectrum and do not track the phase of an individual wave. They use the wave action balance equation or the energy balance equation to determine the statistical properties of a wave field, such as the mean wave height, the mean wave period, and directionality \cite{Porcile2024, Booij1999}. These models have advantages at larger scales than phase-resolving models, including regional and global wave forecasts and long-term climate studies \cite{Sato2023}. Phase-resolving wave models solve variants of the Navier-Stokes, Boussinesq, or nonlinear shallow water equations to describe the elevation and flow fields of the sea surface \cite{Rogers2018, Malej2024, Chondros2024}. They can resolve details about wave behavior, including diffraction, reflection, nonlinear shoaling, wave breaking, and interaction between waves and structures, and are used at smaller scales for coastal engineering like harbor design, wave energy converter optimization, and nearshore circulation \cite{Kim2023, Xu2024, Iphineni2024}.

Phase-averaged spectral wave models evolved into three generations, each representing a new paradigm for modeling and implementing physical processes \cite{Chalikov2023,WAMDIG1988}. The first generation, developed in the 1960s and 1970s, used a simple representation of wave generation by wind and energy dissipation and maintained a specified universal wave spectrum shape \cite{Reidulff2023}. In first-generation models, wind input forced wave growth directly until saturation. At this point, excess energy from saturated waves would dissipate artificially to maintain the shape of the spectral distribution. As such, first-generation models had a fundamental limitation based on their failure to model the complex energy exchange between various frequency components; thus, they produced large errors in modeling swell propagation and spectral distribution changes over time.

The second generation of 1980s models better represented nonlinear wave interactions. It used a simplified four-wave resonance process representation via Hasselmann’s kinetic equation \cite{Hasselmann1985, Chalikov2023}. The wave spectrum evolution was freer than in first-generation models; however, due to the need for computational efficiency, the spectral distribution shape was constrained. The WAM model (an early Wave Model version), a significant advancement from earlier models, had limitations including an incomplete nonlinear energy transfer representation and the need for universal empirical parameters.

The first generation of spectral wave models emerged in the early 1960s, the second in the early 1970s, and the third since the late 1980s, still evolving today \cite{WAMDIG1988,Porcile2024}. They are the best phase-averaged models available, allowing spectral shape determination by solving the wave action balance equation with few assumptions about the spectral distribution shape \cite{Booij1999}. Third-generation models incorporate all physical processes in the energy or action balance equation affecting the wave field, including wind input, nonlinear wave-wave interaction through the discrete interaction approximation (DIA) or a more sophisticated scheme, white capping, bottom friction, depth-induced breaking, and wave-current interaction \cite{Booij1999}. Third-generation models include SWAN \cite{Booij1999}, WAVEWATCH III \cite{Tolman2009}, WAM CYCLE 4 AND LATER VERSIONS \cite{WAMDIG1988},MIKE 21 SW \cite{Sorensen2004}, and TOMAWAC \cite{Benoit2013}. Recent developments of third-generation models aim to improve source term parameterization, notably the ST6 physics package for better wind input and dissipation formulations \cite{Amarouche2023,Rogers2018}, unstructured grids for complex coastal geometries \cite{Roland2012,Zijlema2010}, and new numerical schemes to eliminate computational errors like the ``garden sprinkler" error \cite{Tolman2002}.

Phase-resolving models have evolved significantly. The latest Boussinesq-type models use advanced higher-order dispersion characteristics and better nonlinearity representations. Many recent versions employ advanced wave-breaking schemes \cite{Wei1995,Madsen1992,Tissier2012}. Nonhydrostatic models can model highly nonlinear waves in complex bathymetric and geometric configurations, and fully nonlinear potential flow models (like FUNWAVE \cite{Kirby1998}, COULWAVE, SWASH \cite{Zijlema2011}, and NHWAVE) offer more capability than Boussinesq-type models. GPU acceleration has enabled phase-resolving models to solve problems at rates comparable to or exceeding phase-averaging models \cite{Gao2024}. Adaptive mesh refinement and hybrid approaches combining phase-resolving and phase-averaging strengths are now available.

Recent developments have improved our ability to numerically model waves, but significant challenges remain. A major challenge is accurate numerical representations of wave dynamics, specifically whitecapping dissipation and depth-induced breaking. These processes involve complex turbulent motions and have not been accurately modeled using current models \cite{Ardhuin2010}. Another challenge lies in accurately representing the effects of the wave-current interaction in areas with strong tidal currents, such as ocean eddies, which require sophisticated methods to combine wave models and tidal/ocean eddy models \cite{Ardhuin2017,Wolf2009}. Another challenge is developing numerical models to study Earth's systems by integrating wave models with atmospheric, oceanic circulation, and sediment transport models to create fully coupled Earth system models. This requires solutions to numerical stability and computational efficiency problems and consistent interface process representation between models \cite{Warner2008,Warner2010,Chen2013}. Climate change has impacted wave climates globally, including changes in wave height, period, and directional distribution. To simulate climate change impacts on wave climates globally over long periods (i.e., decades) with high spatial resolutions, large computing power will be required, pushing the limits of current capabilities \cite{Fan2012,CasasPrat2024}. Finally, while advancements have been made in wave forecasting models, these models must incorporate satellite altimeter data, Synthetic Aperture Radar (SAR) observations, and in-situ measurement data to increase forecast accuracy. Advancing data assimilation to effectively combine model output with observational data is necessary \cite{pokhrel2024machine}.

This review considers the current state of numerical wave modeling for coastal and oceanographic purposes. It focuses on the theoretical foundations, numerical techniques, performance, verification, and applications of phase-average and phase-resolving models. The survey aims to (1) provide a systematic review of the mathematical formulations, governing equations, and numerical techniques in modern wave models; (2) critically assess the performance, limitations, and applications of various model classes; and (3) discuss current problems and future research directions in numerical wave modeling. Unlike previous surveys emphasizing certain models or applications, this approach covers the complete range of wave models from global spectral to high-resolution phase-resolving and their applications in operational forecasting, coastal engineering, climate research, and event prediction.

This paper reviews numerical wave modeling approaches to provide a resource for (1) researchers developing new model capabilities, (2) engineers selecting and applying models to practical problems, (3) forecasters needing models to improve predictions, and (4) policymakers seeking scientific guidance for coastal management and climate adaptation decisions.

\section{Review Methodology}
This section summarizes the method for reviewing numerical wave models for coastal and oceanographic purposes. The methodology aspects include literature research, the selection criteria for including and rejecting studies, identifying literature gaps in wave model reviews, and the importance of this review.

\subsection{Literature Search Strategy}
This paper searched the current literature using various scientific databases, including Web of Science, Scopus, Google Scholar, IEEE Xplore, and specific academic repositories (i.e., arXiv, Institutional Archives). The search terms consisted of keyword combinations (e.g., ``numerical wave models," ``spectral wave models," ``phase averaged models," ``phase resolving models," ``Boussinesq equations," ``SWAN," ``WAVEWATCH III," ``coastal wave modeling," ``wave-current interactions," ``model validation," and ``operational forecasting"). In addition to keyword searches, all key reviews and seminal articles have been examined to determine other relevant articles through backward citation tracking. Furthermore, to track recent developments based upon their original contributions, forward citation analysis has been conducted.

Technical documentation, user manuals, and validation reports for operational wave models were obtained from authoritative sources, including the National Oceanic and Atmospheric Administration (NOAA), the European Center for Medium-Range Weather Forecasts (ECMWF), Deltares, the DHI Group, and academic research centers. Conference proceedings from major forums such as the International Conference on Coastal Engineering (ICCE), Coastal Dynamics, and Ocean Sciences Meeting were also reviewed to capture cutting-edge research and emerging methodologies.

\subsection{Inclusion and Exclusion Criteria}
For this review on wave modeling, inclusion criteria included: (1) Peer-reviewed articles, conference presentations, or technical reports on wave models or their applications; (2) Articles on the mathematics and numerical methods in wave models, their performance assessment, or model comparisons; (3) Articles on coastal/oceanographic applications like forecasting, climate studies, coastal engineering, and extreme event predictions; (4) Articles presenting empirical, field, or experimental data for model validation; and (5) Articles on coupling wave models with others, data assimilation techniques, and new methodologies.

The review excluded studies that (1) focused only on theoretical wave physics without numerical models; (2) covered niche applications irrelevant to general coastal and oceanographic contexts; (3) presented unvalidated preliminary results; (4) were in non-peer-reviewed outlets, unless from recognized operational centers; or (5) described solely wave-measuring instruments without numerical model connections. However, foundational theoretical works and seminal articles were included if they provided context for current numerical approaches.

\subsection{Gaps in Existing Review Literature}

Despite multiple reviews of wave modeling over the last twenty years, important gaps still exist, justifying this study. The review by Cavaleri et al. \cite{Cavaleri2007} provided a detailed evaluation of wave modeling as of 2007, discussing the strengths and weaknesses of spectral wave models and highlighting critical research areas. Since then, there have been notable advancements, including new parameterizations of the source term like the ST6 physics package \cite{Rogers2018}, the introduction of unstructured grid capabilities for intricate coastal areas \cite{Zijlema2010,Roland2012}, and the use of GPU-accelerated computing for phase-resolving models. Another review by Cavaleri et al. \cite{Cavaleri2020} covered some of these updates but continued to concentrate mainly on spectral wave models without thoroughly addressing phase-resolution methods.

The review by Thomas and Dwarakish \cite{Thomas2015} provided a useful overview of numerical wave modeling but lacked depth in mathematical formulations and offered a limited discussion of recent advances in model physics and numerical techniques. Lavidas and Venugopal \cite{Lavidas2018} conducted a thorough review of wave model applications on European coastlines, with an emphasis on the assessment of wave energy resources, but their scope was geographically constrained and did not address global-scale modeling or recent developments in the prediction of tropical cyclone waves. The review by Kirby \cite{Kirby2016} offered an excellent treatment of Boussinesq-type models and their applications on a wide range of scales but focused exclusively on phase-resolution approaches without comparative analysis of phase-averaged models or a discussion of appropriate model selection criteria for different applications.

More recent reviews have addressed specific aspects of wave modeling, such as wind-wave climate changes and their impacts \cite{CasasPrat2024}, tropical cyclone wave modeling \cite{Yao2025}, and coastal flooding due to extreme events \cite{Karjoun2023}, but a comprehensive, integrated review spanning both phase-averaged and phase-resolving models with detailed treatment of mathematical formulations, numerical techniques, validation methodologies, and diverse applications has been lacking. Furthermore, existing reviews have provided limited guidance on model selection criteria, computational resource requirements, and trade-offs between accuracy and efficiency for different application contexts. The rapid evolution of computational capabilities, the emergence of hybrid modeling approaches, and growing interest in model coupling for integrated coastal system prediction create a need for an updated, comprehensive assessment that synthesizes recent advances and provides practical guidance for researchers, engineers, and operational forecasters.

\subsection{Motivation and Objectives}

This review is motivated by factors highlighting the importance of accurate wave modeling in modern coastal and oceanographic research. Climate change is altering global wave patterns, with implications for coastal erosion, flooding risk, and marine ecosystem dynamics \cite{CasasPrat2024,Fan2012}. Coastal populations continue to grow, with increasing exposure to wave-related hazards and rising economic stakes associated with coastal infrastructure and maritime operations. The recent catastrophic impact of extreme events including devastating hurricanes, typhoons, and storm surge events has emphasized the pressing need for effective wave forecasting techniques to support emergency management and long-term adaptation considerations \cite{Irham2024,Bertin2015}. At the same time, the growth of renewable energy systems offshore, particularly wave energy conversion production and offshore wind power generation, will require accurate assessment of wave energy resources on the one hand and operational forecasting on the other.

The opportunities and challenges associated with modern wave modeling have been enhanced by advances in technology. The rapid advancement of computer power has made high-resolution simulations impossible formerly. Satellite remote sensing yields vast new areas of coverage in space and time with respect to ocean wave conditions. A growing number of wave forecasting systems operating in the world require rigorous intercomparison and validation so that reliable methods are identified and best practices developed. New methods like machine learning and data assimilation will enhance model accuracy and computation times \cite{schmidt2024forecasting}. However, their evaluation and integration with physics-based techniques require consideration.

This review has three fundamental goals. The first is to give a complete and comprehensive overview of the current mathematical models, governing equations, and numerical methods used in the modeling of waves; these include phase-averaged spectral models and phase-resolving models. We have provided a comprehensive look at how the models treat their source terms, how they discretize the governing equations, and what type of computational algorithms are utilized by the models to determine the model's ability to represent the physics of the system and its applicability. The second goal is to be able to provide critical assessments of the capabilities and limitations of each model class to enable researchers to make informed decisions about which model is most applicable to a particular study or problem. In addition, we hope to provide practical guidance for researchers who select between the various model classes, based on the spatial scales of the study, the temporal resolutions required, the physical processes of interest, and the computational resources available. The third goal is to identify the areas of ongoing research in numerical wave modeling (i.e., issues with the parameterization of unresolved physics, model coupling strategies, validation methodologies, etc.) and to outline potential future research directions in this area.

We anticipate that this review will provide a much-needed resource for many stakeholders in the field. Researchers developing new modeling capabilities will be particularly interested in our review of the current state-of-the-art in both the mathematical formulation and numerical solution of wave models. They will also be able to leverage our identification of gaps in our understanding of the physics represented in current models and our discussion of the research priorities necessary to address those gaps. In addition, engineers and practitioners who apply models to coastal design, coastal hazard assessment, or real-time coastal forecasting will be able to use our review as a guide to select a model that is suitable for their needs and will be able to learn about how to validate models and interpret the results produced by those models. National meteorological and oceanographic centers that conduct real-time coastal forecasting operations will be able to use our review to compare the relative performance of different models and to explore ways to improve their predictive capabilities using new techniques. Finally, policymakers and coastal managers who need to rely on the scientific community to inform their adaptation planning and risk assessment activities will find our review to be a useful synopsis of the capabilities and limitations of models relevant to decision-making.

\subsection{Organization of the Review}

This review systematically progresses from theory to practical implementation. Section 2 outlines the review methodology, including the literature search, inclusion/exclusion criteria, literature gaps, and motivation for this assessment. Section 3 examines the fundamentals of wave theory, model classification, and spectral model evolution with recent advances such as ST6 physics, unstructured grids, and GPU acceleration. Section 4 details mathematical formulations for both phase-averaged and phase-resolving models, covering wave action balance, source term parameterizations, and phase-resolving formulations. Section 5 discusses numerical techniques in wave models, focusing on spatial discretization, temporal integration, boundary conditions, and recent advancements in computational efficiency. Section 6 offers a comparative analysis of operational wave models, evaluating performance, application contexts, validation, and coupled modeling considerations. Section 7 highlights challenges and future research, examining gaps in physical process representation, computational challenges, data assimilation, and emerging research like AI integration and impacts of climate change. Section 8 summarizes the findings, the latest approaches, and the model selection recommendations.

\section{Wave Theory, Model Classification, Selection, and Validation}

\subsection{Fundamental Wave Theory}

Numerical wave models are based on theoretical formulations that describe ocean wave dynamics across different regimes of water depth and wave amplitude. The appropriate theoretical framework depends on two dimensionless parameters: relative water depth $kh$ (where $k$ is the wavenumber and $h$ is the water depth) and wave steepness $ak$ (where $a$ is the wave amplitude) \cite{Mei1983,Dean1991}. Linear wave theory applies when wave steepness is small ($ak \ll 1$), while nonlinear theories are required for steeper waves or shallow water conditions \cite{Komen1994,Holthuijsen2007}.

The dispersion relation for linear waves relates angular frequency $\omega$ to wavenumber $k$ and water depth $h$:

\begin{equation}
\omega^2 = gk\tanh(kh),
\label{eq:dispersion}
\end{equation}
where $g$ is gravitational acceleration. This relation yields the phase velocity $c = \omega/k$ and group velocity $c_g = \partial\omega/\partial k = (c/2)[1 + 2kh/\sinh(2kh)]$. In deep water ($kh > \pi$), Equation \ref{eq:dispersion} simplifies to $\omega^2 = gk$, and $c_g = c/2$, indicating that wave energy propagates at half the phase velocity. In shallow water ($kh < \pi/10$), the relation becomes $\omega^2 = gk^2h$, yielding $c_g = c = \sqrt{gh}$, where energy and phase propagate at the same speed.

Wave energy density per unit horizontal area is given by:

\begin{equation}
E = \frac{1}{2}\rho g a^2 = \frac{1}{8}\rho g H^2,
\label{eq:energy}
\end{equation}
where $\rho$ is water density and $H = 2a$ is wave height. The energy flux (power per unit crest width) is $F = E c_g$. In the presence of ambient currents $\mathbf{U}$, wave action density $A = E/\sigma$ (where $\sigma = \omega - \mathbf{k} \cdot \mathbf{U}$ is the intrinsic frequency) is conserved rather than energy \cite{Bretherton1968,Komen1994}. This conservation principle forms the basis for phase-averaged spectral wave models, as detailed in Section IV.

When wave amplitude increases or water depth decreases, nonlinear effects become significant. Stokes' expansion provides higher-order corrections to linear theory, valid for waves in intermediate to deep water with moderate steepness \cite{Stokes1847,Fenton1985}. For shallow water waves with long wavelengths relative to depth, cnoidal wave theory based on elliptic functions provides accurate descriptions \cite{Korteweg1895,Fenton1999,Isobe1985}. The solitary wave, representing the limiting case of cnoidal waves as wavelength approaches infinity, is described by:

\begin{equation}
\eta(x,t) = H \operatorname{sech}^2\left[\sqrt{\frac{3H}{4h^3}}(x - ct)\right],
\label{eq:solitary}
\end{equation}
where $\eta$ is surface elevation and $c = \sqrt{g(h+H)}$ is the wave celerity. These nonlinear theories are essential for phase-resolving models that simulate detailed wave dynamics in coastal regions.

\subsection{Classification of Numerical Wave Models}

Numerical wave models are classified based on their treatment of phase information, governing equations, and computational approach. The fundamental distinction separates phase-averaged spectral models from phase-resolving deterministic models.

\subsubsection{Phase-Averaged vs. Phase-Resolving Models}

Phase-averaged models represent the sea state statistically through the wave energy or action spectrum $E(f,\theta)$ or $N(f,\theta) = E(f,\theta)/\sigma$, where $f$ is frequency and $\theta$ is direction. These models solve the wave action balance equation, which describes the evolution of wave action density in space and time due to propagation, refraction, and various physical processes including wind input, nonlinear wave-wave interactions, whitecapping dissipation, bottom friction, and depth-induced breaking \cite{Hasselmann1973,Komen1994}. The detailed mathematical formulation of the wave action balance equation and its source terms is presented in Section IV. Phase-averaged models are computationally efficient, with typical computational complexity of $O(N_x N_y N_f N_\theta)$ for $N_x \times N_y$ spatial grid points and $N_f \times N_\theta$ spectral discretization, enabling simulations over large domains and long time periods \cite{Cavaleri2007,Cavaleri2020}.

Phase-resolving models directly compute the temporal and spatial evolution of sea surface elevation $\eta(\mathbf{x},t)$ and associated velocity fields by solving variants of the Navier-Stokes equations, Boussinesq-type equations, or nonlinear shallow water equations \cite{Peregrine1967,Kirby2016}. The governing equations for these models, including various formulations of Boussinesq equations, mild-slope equations, and non-hydrostatic models, are presented in detail in Section IV. Phase-resolving models preserve phase information and explicitly capture wave diffraction, reflection, nonlinear shoaling, wave breaking, and wave-structure interactions. However, they require significantly higher computational resources, with complexity typically $O(N_x N_y N_t)$ for $N_t$ time steps, limiting applications to smaller spatial domains and shorter simulation periods \cite{Kirby2016,Coulaud2025}.

Table \ref{tab:model_comparison} summarizes the key differences between phase-averaged and phase-resolving models.

\begin{table}[!t]
\caption{Comparison of Phase-Averaged and Phase-Resolving Wave Models}
\label{tab:model_comparison}
\centering
\begin{tabular}{p{2.5cm}p{2.5cm}p{2.5cm}}
\toprule
\textbf{Feature} & \textbf{Phase-Averaged Models} & \textbf{Phase-Resolving Models} \\
\midrule
Governing equation & Wave action balance equation & Navier-Stokes, Boussinesq, or mild-slope equations \\
\midrule
Primary variable & Wave action spectrum $N(f,\theta)$ & Surface elevation $\eta(\mathbf{x},t)$ \\
\midrule
Phase information & Not preserved & Fully preserved \\
\midrule
Computational complexity & $O(N_x N_y N_f N_\theta)$ & $O(N_x N_y N_t)$ \\
\midrule
Typical spatial scale & Regional to global (10$^2$--10$^7$ km$^2$) & Local to regional (10$^{-2}$--10$^2$ km$^2$) \\
\midrule
Typical temporal scale & Days to decades & Minutes to days \\
\midrule
Grid resolution & 1--50 km & 1--100 m \\
\midrule
Primary applications & Operational forecasting, climate studies, wave energy assessment & Harbor design, coastal structures, detailed nearshore processes \\
\midrule
Representative models & SWAN, WAVEWATCH III, WAM, MIKE 21 SW, TOMAWAC & FUNWAVE, COULWAVE, SWASH, NHWAVE, BOSZ \\
\bottomrule
\end{tabular}
\end{table}

\subsubsection{Generations of Wave Models}

Spectral wave models have evolved through three distinct generations, characterized by progressively more complete representations of physical processes and reduced empirical constraints \cite{WAMDIG1988,Cavaleri2007}.

\textit{First-Generation Models:} Developed in the 1960s and 1970s, first-generation models employed highly simplified parameterizations in which the wave spectrum was constrained to maintain a prescribed universal shape \cite{Pierson1964,Barnett1968}. Wave growth was represented through direct wind forcing until reaching a saturation limit, at which point energy was artificially dissipated to preserve the spectral form. Nonlinear wave-wave interactions were not explicitly computed. These models were limited by their inability to represent complex energy transfer mechanisms among frequency components, resulting in significant errors in swell propagation and spectral evolution.

\textbf{Second-Generation Models:} Emerging in the 1980s, second-generation models incorporated simplified representations of nonlinear interactions and utilized variable wind fields \cite{Gunther1979,Hasselmann1985}. The wave spectrum was permitted to evolve more freely than in first-generation models, though certain constraints on spectral shape remained to maintain computational feasibility. Nonlinear energy transfer was approximated through simplified parameterizations rather than explicit computation. Examples include the Hybrid Parametrical (HYPA) model and the Spectral Ocean Wave Model (SOWM). While representing significant advances, these models remained limited by incomplete treatment of nonlinear energy transfer.

\textbf{Third-Generation Models:} Introduced in the late 1980s, third-generation models explicitly solve the wave action balance equation without imposing constraints on spectral shape \cite{WAMDIG1988,Komen1994}. All physical processes—wind input, nonlinear wave-wave interactions, whitecapping dissipation, bottom friction, and depth-induced breaking—are represented through explicit parameterizations (detailed in Section IV). Nonlinear four-wave interactions are typically computed using the Discrete Interaction Approximation (DIA) \cite{Hasselmann1985} or more advanced schemes such as the Generalized Multiple DIA (GMD) \cite{Tolman2013}. The DIA approximates the full Boltzmann integral for resonant four-wave interactions using a limited set of representative quadruplets, reducing computational cost from $O(N_f^7)$ for the exact integral to $O(N_f^3)$ for the DIA.

\textbf{Critical Analysis of DIA:} While computationally essential, the DIA introduces unquantified errors into the spectral energy balance. The approximation relies on an arbitrary selection of a small number of quadruplets (typically 4-6) to represent the full continuum of nonlinear interactions. This simplification can lead to inaccuracies in the spectral shape evolution, particularly in complex sea states with bimodal spectra or rapidly changing wind conditions. There is no systematic method to assess the error introduced by the DIA for a given application, and its performance can be sensitive to the specific set of quadruplets used in the model implementation. Consequently, while DIA enables the feasibility of third-generation models, it remains a significant source of uncertainty in the representation of nonlinear wave physics.

Prominent third-generation models include
\begin{itemize}
\item \textbf{WAM (Wave Model):} Developed by the WAMDIG group, WAM was the first operational third-generation model \cite{WAMDIG1988,Komen1994}. Recent versions include WAM Cycle 6 and WAM Cycle 7, which incorporate improved physics packages and unstructured grid capabilities \cite{Ricker2025}.

\item \textbf{WAVEWATCH III:} Developed at NOAA/NCEP, WAVEWATCH III features multiple source term packages, unstructured grid capabilities, and efficient parallelization \cite{Tolman2009,WW3DG2019}. The model supports various physics packages including ST2, ST4, and ST6, with ST6 representing observation-based source terms that provide improved accuracy in extreme conditions \cite{Liu2019,Roh2023}.

\item \textbf{SWAN (Simulating Waves Nearshore):} Designed specifically for coastal applications, SWAN includes enhanced representations of shallow water processes such as depth-induced breaking, bottom friction, and wave-current interaction \cite{Booij1999,Zijlema2010}. SWAN operates on structured, curvilinear, and unstructured grids, with recent versions incorporating the ST6 physics package \cite{Day2022}.

\item \textbf{MIKE 21 SW:} A commercial spectral wave model developed by DHI, MIKE 21 SW includes flexible mesh capabilities and integration with other coastal modeling components \cite{Sorensen2004}.

\item \textbf{TOMAWAC:} Part of the open-source TELEMAC-MASCARET system, TOMAWAC operates on unstructured grids and includes coupling capabilities with hydrodynamic and sediment transport models \cite{Benoit2013}.
\end{itemize}

\textbf{Recent Advances in Third-Generation Models:} Significant developments since 2010 have enhanced the capabilities of third-generation models:

\begin{itemize}
\item \textbf{Improved Source Term Parameterizations:} The ST6 physics package, based on extensive field observations, provides enhanced wind input and whitecapping dissipation formulations that improve accuracy in extreme conditions, including tropical cyclones \cite{Liu2019,Rogers2012,Fernandez2021}. Comparative studies demonstrate that ST6 reduces significant wave height errors by 15--25\% relative to previous parameterizations in hurricane conditions \cite{Roh2023,Day2022,Amarouche2023}.

\item \textbf{Unstructured Grid Capabilities:} Implementation of unstructured triangular meshes enables efficient resolution of complex coastal geometries with local grid refinement \cite{Zijlema2010,Roland2012,Qi2009}. Unstructured grids reduce computational cost by factors of 5--10 compared to structured grids for equivalent accuracy in geometrically complex domains \cite{Wu2018,Feng2016}.

\item \textbf{Advanced Numerical Schemes:} Improved advection schemes reduce numerical artifacts such as the garden sprinkler effect (artificial directional spreading) and enhance stability for large time steps \cite{Tolman2002,Roland2012}. Higher-order schemes including ULTIMATE QUICKEST and third-order upwind methods are now standard options.

\item \textbf{High-Performance Computing:} GPU acceleration and massively parallel implementations enable near-real-time global wave forecasting at high resolution \cite{Yuan2024}. GPU-accelerated versions of WAM achieve speedups of 20--50$\times$ relative to CPU implementations, enabling operational forecasts at 1/12$^\circ$ global resolution \cite{Yuan2024}.

\item \textbf{Coupled Modeling:} Tight coupling with atmospheric, oceanic circulation, and sediment transport models enables simulation of complex interactions in coastal and estuarine environments \cite{Warner2010,Feng2016}. Two-way wave-current coupling accounts for radiation stress gradients, enhanced bottom friction, and Doppler shifting effects.
\end{itemize}

\begin{figure*}[!t]
\centering
\includegraphics[scale=0.25]{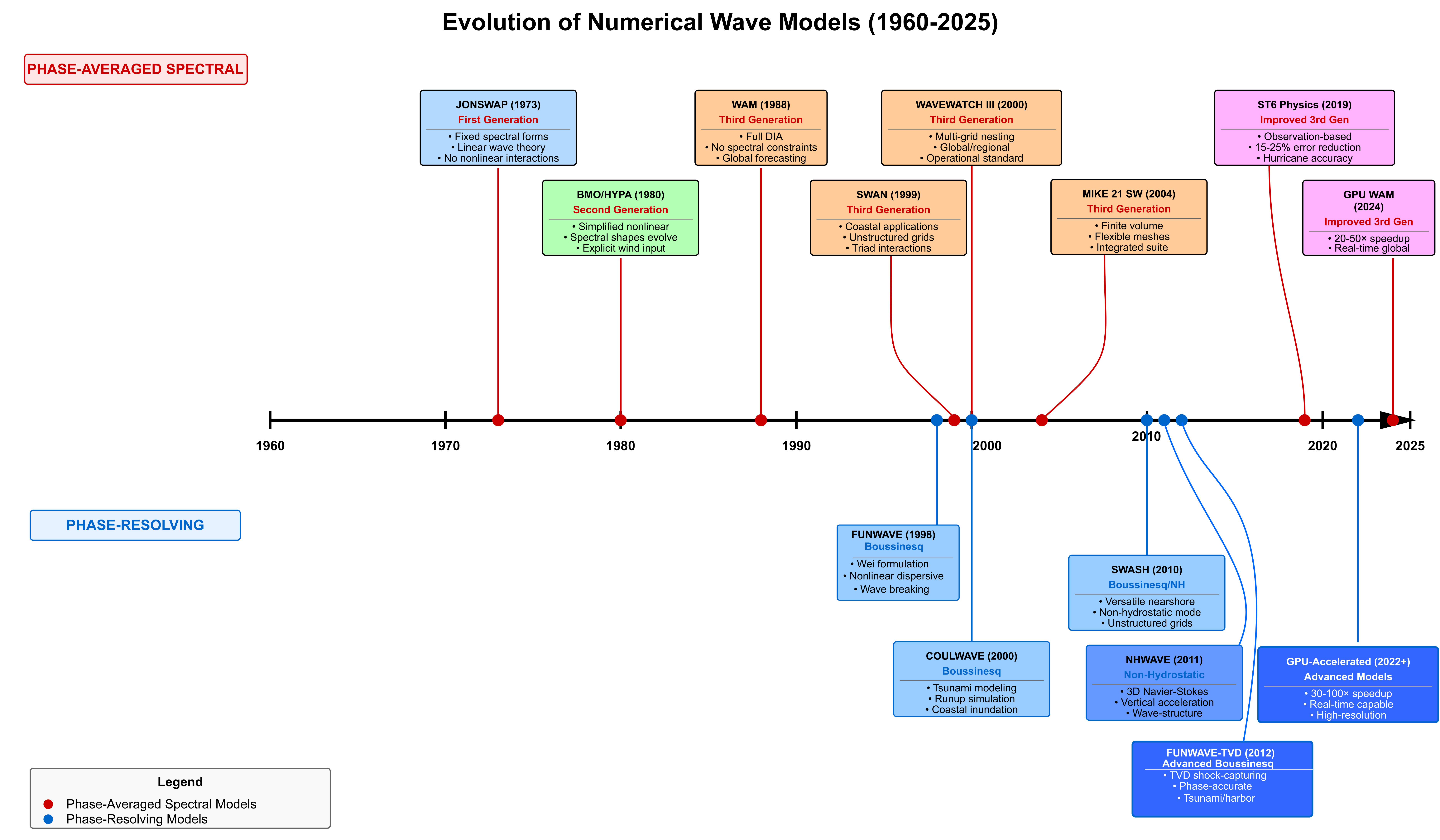}
\caption{Evolution of numerical wave models from 1960 to 2025, showing parallel development of phase-averaged spectral models (top track) and phase-resolving models (bottom track)}
\label{fig:model_evolution}
\end{figure*}

Fig. \ref{fig:model_evolution} illustrates the evolution of wave models through these generations, highlighting the increasing complexity and physical realism of successive generations.

\subsubsection{Phase-Resolving Model Development}

Phase-resolving models have similarly evolved, with modern Boussinesq-type models incorporating higher-order dispersion, improved nonlinearity, and advanced breaking schemes \cite{Kirby2016}. Extended Boussinesq formulations achieve accurate dispersion characteristics from deep to shallow water through additional dispersive terms \cite{Madsen1992,Nwogu1993}. Hybrid schemes combining finite-difference and finite-volume methods enable robust shock-capturing for wave breaking \cite{Tonelli2009,Shi2012}. The mathematical formulations of these models are presented in Section IV.

Representative phase-resolving models include:

\begin{itemize}
\item \textbf{FUNWAVE:} A fully nonlinear Boussinesq wave model with shock-capturing breaking schemes and wetting-drying capabilities \cite{Wei1995,Kirby1998,Shi2012}.

\item \textbf{SWASH (Simulating WAves till SHore):} A non-hydrostatic shallow water model that solves the full shallow water equations without Boussinesq approximations, enabling simulation from deep to very shallow water \cite{Zijlema2011}.

\item \textbf{NHWAVE:} A three-dimensional non-hydrostatic model for wave-structure-sediment interaction \cite{Ma2012}.

\item \textbf{BOSZ:} A Boussinesq model with optimized dispersion and shoaling characteristics \cite{Roeber2010}.
\end{itemize}

Recent advances include GPU acceleration, which reduces computational time by factors of 50--100 \cite{Fang2022,Hwang2025}, and adaptive mesh refinement for efficient resolution of localized features \cite{Popinet2015}. Comparative studies of multiple Boussinesq formulations demonstrate that model selection significantly affects accuracy for specific applications, with no single formulation optimal for all conditions \cite{Coulaud2025}.

\subsection{Spectral Representation}

The wave spectrum $E(f,\theta)$ represents the distribution of wave energy density over frequency $f$ and direction $\theta$. Integrated spectral parameters include:

\begin{equation}
H_s = 4\sqrt{m_0} \quad \text{where} \quad m_0 = \int_0^\infty \int_0^{2\pi} E(f,\theta) \, d\theta \, df
\label{eq:Hs}
\end{equation}

\begin{equation}
T_p = \frac{1}{f_p} \quad \text{where} \quad f_p = \arg\max_f \int_0^{2\pi} E(f,\theta) \, d\theta
\label{eq:Tp}
\end{equation}

where $H_s$ is significant wave height, $m_0$ is the zeroth moment of the spectrum, $T_p$ is peak period, and $f_p$ is peak frequency.

Spectral discretization typically employs logarithmic frequency spacing to efficiently resolve both high-frequency wind sea and low-frequency swell:

\begin{equation}
f_n = f_{\min} \cdot r^{n-1}, \quad n = 1, 2, \ldots, N_f,
\label{eq:freq_discretization}
\end{equation}
where $r = (f_{\max}/f_{\min})^{1/(N_f-1)}$ is the frequency ratio. Typical values are $f_{\min} = 0.04$ Hz, $f_{\max} = 1.0$ Hz, and $N_f = 25$--40. Directional discretization uses uniform spacing with $N_\theta = 24$--36 directions. Spectral resolution requirements depend on application: operational forecasting typically uses $N_f \times N_\theta = 30 \times 24$, while detailed coastal studies may require $40 \times 36$ or finer \cite{Cavaleri2007,Holthuijsen2007}.

Directional spreading functions describe the distribution of wave energy across different propagation directions. Common formulations include the cosine-power function:

\begin{equation}
D(\theta; \theta_m, s) = C(s) \cos^{2s}\left(\frac{\theta - \theta_m}{2}\right),
\label{eq:directional_spreading}
\end{equation}
where $\theta_m$ is mean direction, $s$ is spreading parameter, and $C(s)$ is a normalization constant. Typical values are $s = 2$--10, with higher values indicating narrower directional spreading. Alternative formulations include wrapped normal and von Mises distributions, which better represent bimodal directional distributions in complex sea states \cite{Mitsuyasu1975,Hasselmann1980}.

\subsection{Hybrid and Coupled Approaches}

Recent developments include hybrid approaches that combine phase-averaged and phase-resolving models to leverage the strengths of both methodologies \cite{Abdolali2025}. One-way coupling uses phase-averaged models to provide boundary conditions for nested phase-resolving domains, enabling efficient simulation from deep water to nearshore \cite{Rusu2013}. Two-way coupling allows feedback from phase-resolving to phase-averaged domains, improving accuracy in regions with strong nonlinear interactions \cite{Abdolali2025}. These hybrid approaches reduce computational cost by factors of 10--100 compared to pure phase-resolving simulations over large domains while preserving detailed nearshore dynamics.

Building on the classification and formulations presented in Section III, this section provides comprehensive guidance for model selection, validation, and application. We present comparative analysis of major operational wave models, focusing on relative advantages, limitations, computational requirements, and application domains. We then address model validation approaches and performance assessment, followed by representative application examples across coastal engineering, operational forecasting, and extreme event prediction. Finally, we discuss coupled modeling considerations for comprehensive coastal and ocean simulations. Rather than repeating formulation details from earlier sections, we emphasize practical considerations for model selection, validation strategies, and real-world applications.

\subsection{Phase-Averaged Spectral Models: Comparative Analysis}

Phase-averaged spectral models are the cornerstone of operational wave forecasting and long-term climate studies, offering a balance between physical realism and computational feasibility. While all major models in this category solve the wave action balance equation, they differ significantly in their numerical approaches, grid handling, and target applications. This section provides a comparative overview of the most prominent models: SWAN, WAVEWATCH III, MIKE 21 SW, WAM, and TOMAWAC. A detailed feature-by-feature comparison is provided in Table \ref{tab:phase_averaged_comparison}.

\subsubsection{SWAN}
SWAN (Simulating Waves Nearshore) is a third-generation wave model, developed at Delft University of Technology, that is specifically designed for coastal regions, estuaries, and lakes. Its primary strength lies in its comprehensive representation of shallow-water physics, including triad wave-wave interactions, depth-induced breaking, and bottom friction. The use of a fully implicit time integration scheme makes it particularly efficient for steady-state simulations, such as those required for coastal design and engineering studies. As an open-source model with a large user community, it is extensively documented and validated for a wide range of nearshore applications.

\subsubsection{WAVEWATCH III}
Developed at NOAA/NCEP, WAVEWATCH III is a third-generation wave model that has become the standard for large-scale ocean forecasting. Its key feature is its flexibility, offering multiple grid options (including regular, unstructured, and multi-grid mosaics) and a modular design that allows for the easy implementation of different physics packages (e.g., ST6 for improved storm performance). WAVEWATCH III is particularly well-suited for global and basin-scale applications, and its advanced capabilities for modeling wave-ice interactions make it the preferred choice for polar regions. Its open-source nature and continuous development have led to its widespread adoption in both operational and research settings.

\subsubsection{MIKE 21 SW}
MIKE 21 SW is a commercial third-generation spectral wave model developed by DHI. It is distinguished by its user-friendly graphical interface and seamless integration with other modules in the MIKE suite, which cover hydrodynamics, sediment transport, and water quality. This makes it a powerful tool for comprehensive coastal and marine engineering projects. The model uses a cell-centered finite volume method on an unstructured mesh, providing excellent flexibility for resolving complex coastlines. While its commercial nature and license fees can be a limitation, the professional support and regular updates make it a reliable choice for engineering consultancies and operational agencies.

\subsubsection{WAM}
WAM (Wave Model) was the first third-generation wave model, developed by the WAMDI Group in the 1980s. It set the standard for operational wave forecasting for many years and remains in use at several meteorological centers. WAM was pioneering in its explicit computation of nonlinear wave-wave interactions using the Discrete Interaction Approximation (DIA). While it has been superseded by WAVEWATCH III in many applications due to its more limited grid options (primarily regular latitude-longitude grids), recent developments have included GPU acceleration, which has significantly enhanced its performance for real-time global forecasting.

\subsubsection{TOMAWAC}
TOMAWAC is a third-generation spectral wave model that is part of the open-source TELEMAC-MASCARET modeling system. Its use of the finite element method on unstructured triangular meshes provides high flexibility for resolving complex geometries. TOMAWAC is tightly coupled with the other modules of the TELEMAC system, such as TELEMAC-2D (hydrodynamics) and SISYPHE (sediment transport), making it a strong choice for research and engineering studies that require a comprehensive, integrated modeling approach. Its adoption is particularly widespread in Europe and within the academic research community.

\subsection{Phase-Resolving Models: Comparative Analysis}

Phase-resolving models are essential for applications where detailed information about individual waves is required, such as wave-structure interaction, harbor resonance, and surf zone dynamics. These models are computationally intensive but provide a high-fidelity representation of wave motion by solving more fundamental governing equations than their phase-averaged counterparts. This section compares several leading phase-resolving models: FUNWAVE-TVD, SWASH, and BOSZ. A detailed feature-by-feature comparison is provided in Table \ref{tab:phase_resolving_comparison}.

\subsubsection{FUNWAVE-TVD}
FUNWAVE-TVD is a Boussinesq-type model that has become a benchmark for modeling tsunami propagation and other coastal hazards. Its key feature is the use of a TVD (Total Variation Diminishing) finite volume scheme, which provides excellent shock-capturing capabilities for simulating wave breaking. The model supports both Cartesian and spherical coordinates and includes a two-way multi-grid nesting feature that allows for efficient multi-scale simulations. With recent GPU acceleration, FUNWAVE-TVD has become a powerful and widely used open-source tool for a diverse range of coastal applications.

\subsubsection{SWASH}
SWASH (Simulating WAves till SHore) is a non-hydrostatic wave model that solves the nonlinear shallow water equations. Its non-hydrostatic formulation allows it to accurately represent the vertical structure of wave motion, making it particularly well-suited for simulating surf zone dynamics, wave-structure interaction, and coastal hydrodynamics. SWASH is open-source and has an active user community, and its ability to simulate both wave and hydrodynamic processes in a single framework makes it a versatile tool for coastal research.

\subsubsection{BOSZ}
BOSZ is a high-fidelity research model that solves the Reynolds-Averaged Navier-Stokes (RANS) equations using a high-order spectral method. This approach provides a very detailed and accurate representation of wave physics, including turbulence and sediment transport. Due to its extremely high computational cost, BOSZ is primarily used for fundamental research on wave processes, such as the detailed study of wave breaking and its interaction with the seabed. It requires significant expertise to use effectively and is not intended for large-scale or operational applications.

\onecolumn
\begin{longtable}{p{2.5cm} p{4.5cm} p{4.5cm} p{4.5cm}}
\caption{Comparative Analysis of Major Phase-Averaged Spectral Wave Models}
\label{tab:phase_averaged_comparison}\\
\toprule
\textbf{Feature} & \textbf{SWAN} & \textbf{WAVEWATCH III} & \textbf{MIKE 21 SW} \\
\midrule
\endfirsthead
\caption{(Continued) Comparative Analysis of Major Phase-Averaged Spectral Wave Models}\\
\toprule
\textbf{Feature} & \textbf{SWAN} & \textbf{WAVEWATCH III} & \textbf{MIKE 21 SW} \\
\midrule
\endhead
\bottomrule
\endfoot
\endlastfoot

\textbf{Primary Application} & Coastal, nearshore, and inland waters & Global, regional, and basin-scale ocean forecasting & Coastal and offshore engineering with commercial support \\

\textbf{Key Strengths} & 
\begin{itemize}
    \item Comprehensive shallow water physics (triads, breaking, friction)
    \item Fully implicit time integration for efficiency in steady-state cases
    \item Flexible grid options (structured, unstructured)
    \item Open-source with large user community
\end{itemize} & 
\begin{itemize}
    \item Multi-grid (mosaic) capabilities for multi-scale modeling
    \item Modular design for easy implementation of new physics (e.g., ST6)
    \item Advanced options for wave-ice interaction
    \item Highly efficient parallelization for HPC
\end{itemize} & 
\begin{itemize}
    \item Comprehensive GUI for user-friendly setup and analysis
    \item Seamless integration with other MIKE modules (hydrodynamics, sediment)
    \item Cell-centered finite volume method on unstructured mesh
    \item Reliable commercial support and regular updates
\end{itemize} \\

\textbf{Limitations} & 
\begin{itemize}
    \item Limited diffraction representation
    \item Implicit solver can be slow for large, time-dependent simulations
    \item Less suitable for deep ocean applications
\end{itemize} & 
\begin{itemize}
    \item Shallow water physics less comprehensive than SWAN
    \item Steep learning curve due to numerous options
    \item Explicit time-stepping can be restrictive in shallow water
\end{itemize} & 
\begin{itemize}
    \item Commercial license fees can be a barrier
    \item "Black-box" nature limits research flexibility
    \item Smaller user community than open-source alternatives
\end{itemize} \\

\textbf{Numerical Scheme} & Finite difference, finite volume, or finite element with implicit time integration & Primarily finite difference with third-order explicit time integration & Cell-centered finite volume with multi-sequence explicit integration \\

\textbf{Grid Structure} & Structured, curvilinear, and unstructured & Regular, curvilinear, unstructured, and multi-grid (mosaic) & Unstructured triangular mesh \\

\textbf{License} & Open-source (LGPL) & Open-source (Public Domain) & Commercial \\

\bottomrule
\end{longtable}

\begin{longtable}{p{2.5cm} p{4.5cm} p{4.5cm} p{4.5cm}}
\caption{Comparative Analysis of Major Phase-Resolving Wave Models}
\label{tab:phase_resolving_comparison}\\
\toprule
\textbf{Feature} & \textbf{FUNWAVE-TVD} & \textbf{SWASH} & \textbf{BOSZ} \\
\midrule
\endfirsthead
\caption{(Continued) Comparative Analysis of Major Phase-Resolving Wave Models}\\
\toprule
\textbf{Feature} & \textbf{FUNWAVE-TVD} & \textbf{SWASH} & \textbf{BOSZ} \\
\midrule
\endhead
\bottomrule
\endfoot
\endlastfoot

\textbf{Primary Application} & Tsunami modeling, harbor resonance, and nearshore wave processes & Surf zone dynamics, wave-structure interaction, and coastal hydrodynamics & High-fidelity simulation of wave breaking, turbulence, and sediment transport \\

\textbf{Key Strengths} & 
\begin{itemize}
    \item TVD finite volume scheme for robust shock-capturing (wave breaking)
    \item Two-way multi-grid nesting for multi-scale simulations
    \item GPU acceleration for significant speedup
    \item Validated for a wide range of coastal phenomena
\end{itemize} & 
\begin{itemize}
    \item Non-hydrostatic formulation allows for strong vertical accelerations
    \item Multi-layer approach to resolve vertical flow structure
    \item Simulates both wave and hydrodynamic processes
    \item Open-source with an active user community
\end{itemize} & 
\begin{itemize}
    \item High-order spectral method for high accuracy
    \item Resolves turbulence and sediment transport with advanced models
    \item Provides detailed insight into breaking wave physics
    \item Suitable for fundamental research on wave processes
\end{itemize} \\

\textbf{Limitations} & 
\begin{itemize}
    \item Boussinesq formulation has depth limitations (less accurate in deep water)
    \item Can be computationally expensive for large domains
\end{itemize} & 
\begin{itemize}
    \item Computationally demanding, limiting its use to small domains
    \item Can be complex to set up and run
\end{itemize} & 
\begin{itemize}
    \item Extremely computationally intensive
    \item Limited to small-scale research applications
    \item Requires significant expertise to use effectively
\end{itemize} \\

\textbf{Governing Equations} & Boussinesq-type equations (fully nonlinear) & Non-hydrostatic Navier-Stokes equations & Reynolds-Averaged Navier-Stokes (RANS) equations \\

\textbf{Numerical Scheme} & Finite volume with TVD shock-capturing & Finite difference on a staggered grid & High-order spectral method \\

\textbf{License} & Open-source & Open-source & Open-source \\

\bottomrule
\end{longtable}

\twocolumn
\subsection{Quantitative Model Comparison}

Table \ref{tab:model_comparison_quantitative} provides a quantitative comparison of major wave models across key performance and capability metrics. These values are representative and may vary depending on specific implementation, domain characteristics, and computational resources.

\begin{table*}[!t]
\caption{Quantitative Comparison of Major Wave Models}
\label{tab:model_comparison_quantitative}
\centering
\small
\begin{tabular}{p{2cm}p{1.8cm}p{1.5cm}p{1.5cm}p{1.5cm}p{2cm}p{2cm}}
\toprule
\textbf{Model} & \textbf{Typical Domain Size} & \textbf{Grid Resolution} & \textbf{Time Step} & \textbf{Relative Comp. Cost} & \textbf{Parallelization} & \textbf{Primary Depth Range} \\
\midrule
SWAN & 10$^2$--10$^5$ km$^2$ & 0.1--5 km & 10--60 min & 1.0 (baseline) & OpenMP, limited MPI & $h < 200$ m (shallow to intermediate) \\
\midrule
WAVEWATCH III & 10$^4$--10$^8$ km$^2$ & 1--50 km & 5--30 min & 0.5--2.0 & Excellent MPI, GPU & All depths (optimized for deep) \\
\midrule
MIKE 21 SW & 10$^2$--10$^5$ km$^2$ & 0.1--5 km & 10--60 min & 1.0--1.5 & MPI & $h < 500$ m (shallow to intermediate) \\
\midrule
WAM & 10$^4$--10$^8$ km$^2$ & 5--50 km & 10--30 min & 0.8--1.5 & MPI, GPU & $h > 50$ m (intermediate to deep) \\
\midrule
TOMAWAC & 10$^1$--10$^4$ km$^2$ & 0.05--2 km & 10--60 min & 1.2--1.8 & MPI & $h < 200$ m (shallow to intermediate) \\
\midrule
FUNWAVE-TVD & 10$^{-1}$--10$^2$ km$^2$ & 1--50 m & 0.01--0.5 s & 10--100 & MPI, GPU & $h < 100$ m (shallow, $kh < 3$) \\
\midrule
SWASH & 10$^{-1}$--10$^1$ km$^2$ & 1--20 m & 0.01--0.2 s & 20--200 & Limited & $h < 50$ m (shallow to intermediate) \\
\midrule
COULWAVE & 10$^{-1}$--10$^2$ km$^2$ & 2--50 m & 0.01--0.5 s & 15--150 & MPI & $h < 200$ m (intermediate to shallow, $kh < 6$) \\
\bottomrule
\end{tabular}
\end{table*}

The relative computational cost is normalized to SWAN for a representative coastal application (100 km$^2$ domain, 500 m resolution, 24-hour simulation). Phase-resolving models are 10--200 times more expensive due to finer spatial and temporal resolution requirements. Actual costs depend strongly on domain size, resolution, and specific application requirements.

\subsection{Model Selection Guidelines}

Selecting an appropriate wave model requires consideration of multiple factors, including application objectives, domain characteristics, available computational resources, and required output variables. This section provides practical guidelines for model selection.

\subsubsection{Decision Framework}

The primary factors guiding model selection are the spatial scale and domain size of the area of interest. For harbor to local coastal domains smaller than $10\,\text{km}^2$, suitable options include SWAN, TOMAWAC, or phase-resolving models. For regional coastal areas ranging from $10$ to $1{,}000\,\text{km}^2$, SWAN, MIKE 21 SW, and TOMAWAC are recommended. At basin scales of approximately $10^4$ to $10^6\,\text{km}^2$, WAVEWATCH III and WAM are typically employed, while for global domains larger than $10^7\,\text{km}^2$, WAVEWATCH III and WAM remain the preferred modeling tools.

Water depth and associated physics requirements guide the selection of appropriate wave models: in very shallow waters (less than 10 m) with breaking waves, SWAN and other phase-resolving models are recommended; in shallow to intermediate depths (10–200 m), suitable options include SWAN, MIKE 21 SW, TOMAWAC, and WAVEWATCH III; in deep ocean environments (greater than 200 m), WAVEWATCH III and WAM are preferred; and in polar regions with sea ice, WAVEWATCH III is the model of choice.

The required outputs and associated physics are as follows: bulk wave parameters (significant wave height Hs, peak period Tp, and direction) and the full directional spectrum can be provided by any phase-averaged model; phase-resolved surface elevation and velocity fields require the use of phase-resolving models; wave–current interactions can be represented in SWAN, MIKE 21 SW, TOMAWAC, and in phase-resolving models; diffraction around coastal and offshore structures is captured by phase-resolving models, with SWAN offering only limited diffraction capability; and tsunami propagation and inundation analyses are typically conducted using models such as FUNWAVE-TVD, COULWAVE, or SWASH.

Computational resource requirements vary by model and domain size. Desktop workstations are generally sufficient for running SWAN and MIKE 21 SW on small domains. Small clusters with approximately 10–100 cores can efficiently support SWAN, regional WAVEWATCH III, and TOMAWAC simulations. Large high-performance computing (HPC) systems with 100–1000 or more cores are typically required for global WAVEWATCH III, WAM, and FUNWAVE-TVD applications. Additionally, GPU acceleration is available for WAVEWATCH III, WAM, and FUNWAVE-TVD, enabling further performance gains for computationally intensive simulations.

Software and support considerations include the requirement for open-source solutions such as SWAN, WAVEWATCH III, TOMAWAC, FUNWAVE-TVD, SWASH, and COULWAVE, while commercial support is needed for MIKE 21 SW; among these, SWAN and WAVEWATCH III benefit from a large and active user community, and integration capabilities are available through the MIKE suite for MIKE 21 SW and the TELEMAC system for TOMAWAC.

\subsubsection{Application-Specific Recommendations}

\textbf{Coastal Engineering Design:}
For most coastal engineering applications (breakwater design, beach nourishment, coastal structures), SWAN is the preferred choice due to comprehensive shallow water physics, efficient steady-state computation, and extensive validation. For projects requiring diffraction analysis around structures, phase-resolving models (FUNWAVE-TVD, SWASH) should be used for detailed local analysis, potentially nested within SWAN for boundary conditions.

\textbf{Operational Wave Forecasting:}
Global operational forecasting is dominated by WAVEWATCH III due to flexible grid options, efficient parallelization, and operational reliability. Regional coastal forecasting typically uses SWAN or MIKE 21 SW, depending on institutional preferences and computational resources. WAM continues to be used at some centers due to historical continuity and recent GPU acceleration capabilities.

\textbf{Climate Studies and Reanalysis:}
Long-term climate simulations and reanalysis projects typically use WAVEWATCH III or WAM for global domains, with SWAN for regional coastal climate studies. The choice depends on required spatial resolution, available computational resources, and consistency with other climate model components.

\textbf{Tsunami Modeling:}
Tsunami propagation and inundation require phase-resolving models to capture long-wave dynamics, dispersion, and run-up. FUNWAVE-TVD is widely used due to robust TVD numerics, multi-scale nesting capabilities, and extensive validation. COULWAVE and SWASH are alternatives for specific applications. Basin-scale propagation may use WAVEWATCH III or specialized tsunami models, with phase-resolving models for nearshore inundation.

\textbf{Harbor and Port Engineering:}
Harbor resonance, wave agitation, and vessel motion studies require phase-resolving models to capture diffraction, reflection, and resonance phenomena. FUNWAVE-TVD and SWASH are commonly used, with SWAN providing boundary conditions and preliminary analysis. For very complex harbor geometries or wave-structure interactions, CFD models may be necessary.

\textbf{Offshore Engineering:}
Offshore wind farm site assessment, oil platform design, and marine operations typically use WAVEWATCH III for basin-scale wave climate and extreme event analysis. SWAN may be used for nearshore wind farms or when detailed shallow water physics are required. Phase-resolving models are used for detailed wave-structure interaction analysis.

\textbf{Research Applications:}
Research on wave physics, parameterizations, or numerical methods benefits from open-source models with modular design. WAVEWATCH III is preferred for deep-water physics research, SWAN for coastal processes, and FUNWAVE-TVD for phase-resolving nearshore dynamics. The choice depends on specific research objectives and the availability of alternative parameterizations or numerical schemes.

\subsection{Computational Performance Considerations}

Computational efficiency is a critical factor in model selection, particularly for operational forecasting, climate studies, or ensemble simulations. Table \ref{tab:computational_performance} provides representative computational requirements for typical applications.

\begin{table}[!t]
\caption{Representative Computational Requirements for Typical Applications}
\label{tab:computational_performance}
\centering
\small
\begin{tabular}{p{2.5cm}p{2cm}p{1.5cm}p{1.5cm}}
\toprule
\textbf{Application} & \textbf{Recommended Model} & \textbf{Typical Resources} & \textbf{Wall-Clock Time} \\
\midrule
Global forecast (0.5° res, 7-day) & WAVEWATCH III & 100--500 cores & 0.5--2 hours \\
\midrule
Regional coastal (100 km$^2$, 500 m, 24-hour) & SWAN & 1--10 cores & 0.5--2 hours \\
\midrule
Harbor study (1 km$^2$, 10 m, 1-hour) & FUNWAVE-TVD & 10--100 cores & 2--10 hours \\
\midrule
Tsunami basin propagation (10$^6$ km$^2$, 1 km, 6-hour) & WAVEWATCH III & 50--200 cores & 1--4 hours \\
\midrule
Tsunami inundation (10 km$^2$, 5 m, 1-hour) & FUNWAVE-TVD & 50--500 cores & 5--50 hours \\
\midrule
Climate reanalysis (global, 40-year) & WAVEWATCH III & 500--2000 cores & Days to weeks \\
\bottomrule
\end{tabular}
\end{table}

These estimates assume modern HPC systems and efficient parallelization. Actual performance depends on specific hardware, model configuration, and domain characteristics. GPU acceleration can provide a 10--50$\times$ speedup for supported models (WAVEWATCH III, WAM, and FUNWAVE-TVD).

\subsection{Emerging Trends and Future Directions}

Model development continues to advance in several directions that will influence future model selection:

\textbf{Hybrid Approaches:} Coupling phase-averaged and phase-resolving models enables efficient multi-scale simulations, with spectral models providing boundary conditions for phase-resolving models in regions of interest. Automated nesting frameworks reduce user burden and improve consistency.

\textbf{GPU Acceleration:} GPU implementations of spectral models (WAM, WAVEWATCH III) and phase-resolving models (FUNWAVE-TVD) are enabling real-time forecasting and ensemble simulations previously infeasible. This trend will continue as GPU architectures become more prevalent in HPC systems.

\textbf{Machine Learning Integration:} Data-driven parameterizations and surrogate models may supplement or replace traditional physics-based formulations for specific processes, potentially improving accuracy and efficiency. However, physics-based models will remain essential for extrapolation beyond training data.

\textbf{Coupled Earth System Models:} Tighter coupling with atmospheric, oceanic, and ice models within Earth system frameworks requires efficient wave models with consistent numerics and data structures. WAVEWATCH III and WAM are increasingly integrated into coupled systems for climate applications.

\textbf{Unstructured Mesh Advances:} Continued development of unstructured mesh capabilities in spectral models (SWAN, WAVEWATCH III, TOMAWAC) provides increasing flexibility for multi-scale coastal applications, potentially reducing the need for complex nesting configurations.

These trends suggest that model selection will increasingly depend on specific application requirements, available computational resources, and integration with broader modeling frameworks rather than fundamental model capabilities.

\subsection{Model Validation and Performance Assessment}

Validation and verification are essential for establishing the credibility and accuracy of numerical wave models. Validation assesses how well a model represents the real physical system, while verification ensures that the model correctly solves the underlying mathematical equations \cite{Holthuijsen2007}. This subsection provides an overview of validation approaches, data sources, and key metrics used to evaluate wave model performance.

\subsubsection{Validation Data Sources}

Wave model validation relies on diverse observational data sources, each with distinct advantages and limitations. In-situ measurements from wave buoys provide the most reliable point measurements of wave parameters, including full spectral information, and serve as the primary reference for model validation despite limited spatial coverage \cite{Bidlot2002}. Fixed platforms with mounted instruments offer continuous measurements at specific locations, particularly valuable for long-term validation studies \cite{Stopa2016}. Bottom-mounted instruments such as Acoustic Doppler Current Profilers (ADCPs) and pressure sensors measure wave parameters in shallow waters, providing critical data for validating coastal wave models \cite{Work2011}.

Remote sensing technologies extend validation capabilities to broader spatial scales. Satellite altimetry from missions such as Jason-3 and Sentinel-3 measures significant wave height with global coverage, though with limited spatial and temporal resolution \cite{Stopa2016}. Synthetic Aperture Radar (SAR) provides information on wave spectra, particularly for longer wavelengths, useful for validating directional and frequency distributions \cite{Collard2005}. Shore-based high-frequency (HF) radar systems measure wave parameters over coastal areas with better spatial coverage than buoys but lower accuracy \cite{Wyatt2006}. Controlled laboratory experiments in wave tanks and basins provide valuable data for validating specific processes such as wave breaking, nonlinear interactions, and wave-structure interactions under controlled conditions \cite{Mase1992}.

\subsubsection{Statistical Validation Metrics}

Wave model performance is typically quantified using standard statistical metrics. Bias measures systematic error as the mean difference between model predictions and observations. Root Mean Square Error (RMSE) quantifies the average magnitude of errors, while the Scatter Index (SI) normalizes RMSE by the mean observation to facilitate comparison across different conditions. The correlation coefficient ($R$) measures linear association between model and observations, with values ranging from -1 to 1. The coefficient of determination ($R^2$) represents the proportion of observational variance explained by the model \cite{Holthuijsen2007}.

More specialized metrics provide additional insights into model performance. The symmetric slope accounts for errors in both magnitude and phase, while Willmott's Index of Agreement ranges from 0 to 1, with 1 indicating perfect agreement. The Brier Skill Score (BSS) compares model performance against a baseline or reference prediction, with positive values indicating improvement over the baseline \cite{Mentaschi2013}.

Beyond bulk parameters such as significant wave height and peak period, detailed validation can compare full wave spectra using spectral distance parameters like Earth Mover's Distance or Wasserstein distance. Spectral partitioning separates wind sea and swell components, allowing validation of specific wave systems. Directional spread comparison validates the directional distribution of wave energy, critical for applications where wave direction is important \cite{Mentaschi2013}.

\subsubsection{Model Intercomparison and Benchmarking}

Model intercomparison projects provide systematic frameworks for comparing different wave models under identical conditions, helping identify strengths and weaknesses of different approaches. The WISE Group (Wave Modelling Group) has conducted comprehensive reviews and comparisons of wave models, identifying key areas for improvement \cite{Cavaleri2007}. The JCOMM Wave Forecast Verification Project compares operational wave forecasts from different centers against observations, providing valuable information on operational model performance \cite{Bidlot2007}. The World Meteorological Organization has established wave model verification standards to ensure consistency in validation practices across the international community \cite{WMO2016}.

Sensitivity analysis examines how changes in model inputs or parameters affect outputs, providing insights into model behavior and identifying critical factors influencing performance. Parameter sensitivity tests the model's response to variations in physical parameters related to wave generation, dissipation, or nonlinear interactions. Input data sensitivity assesses how uncertainties or errors in wind fields or bathymetry affect model results. Resolution sensitivity examines the impact of spatial, temporal, and spectral resolution on model accuracy and computational efficiency \cite{Tolman1995}.

\section{Mathematical Formulations and Numerical Methods}

Building on the classification framework presented Section III, this section provides detailed mathematical formulations and governing equations for numerical wave models. We begin with the wave action balance equation that governs phase-averaged spectral models, followed by comprehensive descriptions of source term parameterizations representing physical processes. Subsequently, we present the governing equations for phase-resolving models, including Boussinesq-type equations, mild-slope equations, non-hydrostatic formulations, and computational fluid dynamics approaches. These mathematical foundations are essential for understanding model capabilities, limitations, and appropriate applications.

\subsection{Wave Action Balance Equation}

The fundamental equation governing phase-averaged wave models is the action balance equation, which describes the evolution of the wave action density spectrum $N(x,y,\sigma,\theta,t)$ in time and space \cite{Komen1994}:

\begin{equation}
\frac{\partial N}{\partial t} + \nabla_{x} \cdot [(\mathbf{c}_g + \mathbf{U})N] + \frac{\partial c_{\sigma}N}{\partial \sigma} + \frac{\partial c_{\theta}N}{\partial \theta} = \frac{S_{tot}}{\sigma}
\label{eq:action_balance_general}
\end{equation}

where $N = E/\sigma$ is the action density (with $E$ being the energy density and $\sigma$ the relative angular frequency), $\mathbf{c}_g$ is the group velocity vector, $\mathbf{U}$ is the ambient current vector, and $c_{\sigma}$ and $c_{\theta}$ are the propagation velocities in spectral space (frequency and direction, respectively). The term $S_{tot}$ represents the source terms accounting for energy input, dissipation, and redistribution.

The left-hand side of Equation \ref{eq:action_balance_general} represents the kinematic part, describing the propagation of wave action in geographic and spectral spaces. The right-hand side contains the source terms that represent the physical processes affecting the wave spectrum \cite{Booij1999}. In Cartesian coordinates, the action balance equation can be written as:

\begin{equation}
\frac{\partial N}{\partial t} + \frac{\partial c_x N}{\partial x} + \frac{\partial c_y N}{\partial y} + \frac{\partial c_{\sigma} N}{\partial \sigma} + \frac{\partial c_{\theta} N}{\partial \theta} = \frac{S_{tot}}{\sigma}
\label{eq:action_balance_cartesian}
\end{equation}

For large-scale applications, such as global wave modeling, it is more appropriate to use spherical coordinates:

\begin{equation}
\frac{\partial N}{\partial t} + \frac{\partial c_{\lambda} N}{\partial \lambda} + \frac{\partial c_{\phi} N}{\partial \phi} + \frac{\partial c_{\sigma} N}{\partial \sigma} + \frac{\partial c_{\theta} N}{\partial \theta} = \frac{S_{tot}}{\sigma}
\label{eq:action_balance_spherical}
\end{equation}

where $\lambda$ and $\phi$ are longitude and latitude, respectively, and $c_{\lambda}$ and $c_{\phi}$ are the corresponding propagation velocities. The choice of coordinate system affects numerical implementation and accuracy, particularly for global models where spherical geometry must be properly accounted for to avoid singularities at the poles \cite{Tolman2009,WW3DG2019}.

\subsection{Source Terms}

The source/sink term $S_{tot}$ in the action balance equation represents all physical processes that contribute to the generation, dissipation, or redistribution of wave energy. It can be expressed as the sum of several components \cite{Komen1994}:

\begin{equation}
S_{tot} = S_{in} + S_{nl} + S_{ds} + S_{bot} + S_{br} + S_{tr}
\label{eq:source_terms_detailed},
\end{equation}
where:
\begin{itemize}
\item $S_{in}$ is the wind input (energy transfer from wind to waves)
\item $S_{nl}$ is the nonlinear wave-wave interactions
\item $S_{ds}$ is the dissipation due to whitecapping (deep water breaking)
\item $S_{bot}$ is the dissipation due to bottom friction
\item $S_{br}$ is the dissipation due to depth-induced breaking
\item $S_{tr}$ is the energy loss due to wave-obstacle interactions (e.g., vegetation, structures)
\end{itemize}

Different wave models may include additional source terms or use different formulations for each term. The accurate parameterization of these source terms is crucial for the performance of wave models \cite{Ardhuin2010}. The relative importance of each term varies with location and conditions: wind input and whitecapping dominate in deep water, while bottom friction and depth-induced breaking become significant in shallow coastal waters.

\subsubsection{Wind Input}

The wind input term represents the energy transfer from the wind to the waves. It depends on the wind speed, the wave steepness, and the wave-wind angle. Several formulations exist, with many modern models using variants of Janssen's quasi-linear theory \cite{Janssen1991}:

\begin{equation}
S_{in}(\sigma, \theta) = \beta \cdot E(\sigma, \theta),
\label{eq:wind_input}
\end{equation}
where $\beta$ is the exponential growth rate, which depends on the friction velocity, the wave phase speed, and the wave direction relative to the wind direction. Janssen's formulation couples the wave field to the atmospheric boundary layer through the surface roughness, which depends on the wave spectrum itself, creating a feedback mechanism \cite{Janssen1991,Janssen2004}.

Alternative formulations include the ST6 physics package, which is based on extensive field observations and provides improved performance in extreme wind conditions \cite{Rogers2012,Liu2019}. The ST6 wind input formulation accounts for the sheltering effect of longer waves on shorter waves and includes a dependence on wave age (the ratio of wave phase speed to wind speed), which better represents the physics of wind-wave interaction across different sea states.

\subsubsection{Nonlinear Wave-Wave Interactions}

Nonlinear interactions redistribute energy within the wave spectrum and are crucial for the spectral shape evolution. For deep water, the dominant process is the four-wave (quadruplet) interaction, while in shallow water, three-wave (triad) interactions become important \cite{Hasselmann1962,Eldeberky1996}.

The exact computation of nonlinear quadruplet interactions requires evaluation of the Boltzmann integral:

\begin{align}
S_{nl}(\mathbf{k}) &= 
\iiint 
T(\mathbf{k},\mathbf{k}_1,\mathbf{k}_2,\mathbf{k}_3) \,
\delta(\mathbf{k}+\mathbf{k}_1-\mathbf{k}_2-\mathbf{k}_3) \notag\\
&\quad\times 
\delta(\omega+\omega_1-\omega_2-\omega_3) \,
d\mathbf{k}_1\, d\mathbf{k}_2\, d\mathbf{k}_3,
\label{eq:boltzmann_integral}
\end{align}
where $T$ is the interaction coefficient and the delta functions enforce resonance conditions. This computation is extremely complex and computationally intensive, with complexity $O(N_f^7)$ for direct evaluation. Therefore, most operational models use the Discrete Interaction Approximation (DIA) \cite{Hasselmann1985} or other approximations. The DIA approximates the full Boltzmann integral by considering only a limited set of wave number configurations, significantly reducing computational cost to $O(N_f^3)$ while capturing the essential energy transfer characteristics.

More advanced approximations include the Generalized Multiple DIA (GMD) \cite{Tolman2013}, which expands the set of representative quadruplets to improve accuracy while maintaining computational efficiency. Recent developments also include the exact nonlinear (XNL) method \cite{VanVledder2006}, which computes the full Boltzmann integral but remains too expensive for operational applications.

For shallow water, triad interactions transfer energy from lower to higher frequencies through resonant three-wave interactions. The Lumped Triad Approximation (LTA) \cite{Eldeberky1996} is commonly used in coastal models such as SWAN to represent this process.

\subsubsection{Whitecapping Dissipation}

Whitecapping is the process where waves become unstable and break in deep water, dissipating energy. This process depends on wave steepness and is typically parametrized based on the overall steepness of the wave spectrum. A widely used formulation is based on Hasselmann's pulse-based model \cite{Hasselmann1974}:

\begin{equation}
S_{ds}(\sigma, \theta) = -\Gamma \cdot \sigma \cdot \left(\frac{\hat{\sigma}}{\sigma}\right)^n \cdot E(\sigma, \theta)
\label{eq:whitecapping}
\end{equation}

where $\Gamma$ is a coefficient that depends on the overall wave steepness, $\hat{\sigma}$ is a mean frequency, and $n$ is a parameter controlling the dependence on frequency. The coefficient $\Gamma$ is typically expressed as:

\begin{equation}
\Gamma = C_{ds} \left(\frac{\tilde{s}}{\tilde{s}_{PM}}\right)^p
\label{eq:gamma_coefficient}
\end{equation}

where $C_{ds}$ is a calibration coefficient, $\tilde{s}$ is the overall wave steepness, $\tilde{s}_{PM}$ is the Pierson-Moskowitz steepness, and $p$ is an exponent (typically $p = 2$).

The ST6 physics package uses an alternative saturation-based formulation that better represents observations of whitecapping in various sea states \cite{Rogers2012,Liu2019}. This formulation relates dissipation directly to the saturation spectrum and includes a threshold below which dissipation is negligible, better representing the intermittent nature of whitecapping.

\textbf{Critical Analysis of Whitecapping Parameterizations:} Despite these advances, the parameterization of whitecapping dissipation remains one of the most significant sources of uncertainty in wave modeling. The formulations are fundamentally semi-empirical, and their reliance on tunable coefficients raises critical questions about their reliability and physical basis \cite{Ardhuin2010}. The whitecapping dissipation coefficient, $C_{ds}$, is often determined through a trial-and-error process for specific regions and wind products, rather than being derived from first principles. This practice undermines the universality of the models and leads to significant forecast errors when models are applied outside their calibration range.

Recent studies have shown that the optimal $C_{ds}$ value exhibits a compensatory relationship with errors in the forcing wind field, meaning the coefficient is often tuned to mask inaccuracies in the wind input rather than to represent the true physics of wave breaking \cite{sherwood2022modeling}. This co-dependence is particularly problematic under extreme wind conditions, such as in tropical cyclones, where standard-tuned models systematically underpredict significant wave heights by 2--3 meters or more. While the ST6 package has reduced these errors, it has not eliminated them, and the fundamental challenge of developing a physically robust and universally applicable whitecapping parameterization remains unresolved.

\subsubsection{Bottom Friction and Depth-Induced Breaking}

In shallow waters, two additional dissipation processes become important: bottom friction and depth-induced breaking. Bottom friction represents the energy dissipation due to wave-induced oscillatory currents interacting with the seabed. It is typically modeled using a quadratic friction law \cite{Hasselmann1973,Madsen1988}:

\begin{equation}
S_{bot}(\sigma, \theta) = -C_f \frac{\sigma^2}{g^2 \sinh^2(kh)} E(\sigma, \theta)
\label{eq:bottom_friction}
\end{equation}

where $C_f$ is a bottom friction coefficient, which may depend on bottom roughness. Typical values range from $C_f = 0.015$ m$^2$s$^{-3}$ for sandy bottoms to $C_f = 0.067$ m$^2$s$^{-3}$ for very rough bottoms \cite{Hasselmann1973}. Alternative formulations include the JONSWAP formulation \cite{Hasselmann1973} and the Madsen formulation \cite{Madsen1988}, which account for different bottom roughness characteristics.

Depth-induced breaking occurs when waves enter very shallow water and become unstable due to the limited water depth. A common approach is based on the Battjes and Janssen model \cite{Battjes1978}, which assumes that the wave height distribution in shallow water is described by a Rayleigh distribution truncated at a maximum wave height proportional to the water depth:

\begin{equation}
S_{br}(\sigma, \theta) = -\frac{\alpha}{\pi} \frac{Q_b}{T_m} \frac{H_m^2}{16} \frac{E(\sigma,\theta)}{E_{tot}}
\label{eq:depth_breaking}
\end{equation}

where $\alpha$ is a dissipation coefficient (typically $\alpha = 1.0$), $Q_b$ is the fraction of breaking waves, $T_m$ is the mean period, $H_m$ is the maximum wave height (typically $H_m = \gamma h$ with $\gamma \approx 0.73$), and $E_{tot}$ is the total energy density. The fraction of breaking waves $Q_b$ is determined iteratively from the wave height distribution.

Alternative breaking formulations include the extended Battjes-Janssen model \cite{Salmon2015}, which accounts for depth-limited and steepness-limited breaking, and the bore-based model \cite{Thornton1978}, which relates dissipation to the energy flux gradient.

\textbf{Critical Analysis of Depth-Induced Breaking Models:} The widely used Battjes-Janssen formulation relies on an empirical parameter, $\gamma$, to relate the maximum wave height to the water depth. While typically set to 0.73, observations show that $\gamma$ can vary from 0.4 to 1.2 depending on beach slope and incident wave conditions. This reliance on a tunable, non-universal parameter introduces significant uncertainty into surf zone predictions. The model also assumes that all waves break at the same depth-to-height ratio, which is an oversimplification that does not account for the random nature of wave breaking in the surf zone. Consequently, these parameterizations can lead to substantial errors in predicting nearshore wave transformation and runup, particularly on complex or steep bathymetries.

\subsection{Boussinesq Equations for Phase-Resolving Models}

While phase-averaged models focus on the evolution of the wave spectrum, phase-resolving models simulate the actual wave motion, preserving phase information. One important class of phase-resolving models is based on Boussinesq-type equations, which are depth-integrated approximations of the Navier-Stokes equations \cite{Peregrine1967}.

The classical Boussinesq equations are derived by assuming weak nonlinearity ($a/h \ll 1$, where $a$ is the wave amplitude and $h$ is the water depth) and weak dispersion ($h/L \ll 1$, where $L$ is the wavelength). In terms of free surface elevation $\eta$ and depth-averaged horizontal velocity $\mathbf{u}$, they can be written as:

\begin{equation}
\frac{\partial \eta}{\partial t} + \nabla \cdot [(h + \eta)\mathbf{u}] = 0
\label{eq:boussinesq_continuity}
\end{equation}

\begin{equation}
\frac{\partial \mathbf{u}}{\partial t} + (\mathbf{u} \cdot \nabla)\mathbf{u} + g\nabla\eta + \frac{h^2}{3}\nabla(\nabla \cdot \frac{\partial \mathbf{u}}{\partial t}) = 0
\label{eq:boussinesq_momentum}
\end{equation}

The last term in Equation \ref{eq:boussinesq_momentum} represents the dispersive effects due to vertical acceleration. These classical equations are accurate for $kh < 1$ and provide reasonable dispersion characteristics for shallow water waves.

Modern Boussinesq-type models have extended these equations to improve their dispersion characteristics and applicability to deeper water and higher wave nonlinearity \cite{Madsen1992,Wei1995}. Enhanced Boussinesq equations introduce additional dispersive terms or use velocity variables at different vertical positions to achieve improved dispersion relations. For example, the Nwogu formulation \cite{Nwogu1993} uses velocity at an arbitrary vertical level $z_\alpha$ to optimize dispersion:

\begin{align}
\frac{\partial \eta}{\partial t}
&\;+\; \nabla \cdot \!\left[(h+\eta)\mathbf{u}_\alpha\right]  \notag\\
&\;+\; \nabla \cdot \!\left\{(h+\eta)
  \left[(z_\alpha+h)\,\nabla(\nabla\cdot\mathbf{u}_\alpha)\right]\right\}
= 0 ,
\label{eq:nwogu_continuity}
\end{align}

\begin{align}
\frac{\partial \mathbf{u}_\alpha}{\partial t}
&\;+\; (\mathbf{u}_\alpha \cdot \nabla)\mathbf{u}_\alpha
\;+\; g\,\nabla\eta  \notag\\[2pt]
&\;+\; \nabla\!\Big[
   (z_\alpha+h)\,\nabla\!\cdot\!
   \frac{\partial \mathbf{u}_\alpha}{\partial t}
   \Big]                                      \notag\\[2pt]
&\;+\; \nabla\!\Big[
   \tfrac12 (z_\alpha+h)^2\,
   \nabla\!\big(\nabla\!\cdot\!
   \tfrac{\partial \mathbf{u}_\alpha}{\partial t}\big)
   \Big]
= 0 ,
\label{eq:nwogu_momentum}
\end{align}

By choosing $z_\alpha = -0.531h$, the linear dispersion relation is accurate up to $kh \approx 3$, extending the applicability to intermediate water depths \cite{Nwogu1993}.

Fully nonlinear Boussinesq models \cite{Wei1995} remove the weak nonlinearity assumption, allowing simulation of highly nonlinear waves including solitary waves and cnoidal waves. These models typically use higher-order dispersion terms to maintain accuracy in deeper water while capturing strong nonlinear effects in shallow water.

\subsection{Other Governing Equations for Phase-Resolving Models}

Besides Boussinesq-type models, other phase-resolving approaches include:

\subsubsection{Mild-Slope Equation}

The mild-slope equation is a simplified wave equation based on linear wave theory, suitable for modeling wave propagation over slowly varying bathymetry \cite{Berkhoff1972}. It accounts for refraction and diffraction but neglects nonlinear effects. The time-dependent mild-slope equation can be written as:

\begin{equation}
\nabla \cdot (CC_g \nabla \eta) - \frac{\omega^2}{g}\eta = 0
\label{eq:mild_slope}
\end{equation}

where $C$ is the phase velocity and $C_g$ is the group velocity. Extensions include the modified mild-slope equation \cite{Chamberlain1995} and the complementary mild-slope equation \cite{Massel1993}, which improve accuracy for steeper bottom slopes and higher frequencies.

\subsubsection{Non-Hydrostatic Wave Models}

Non-hydrostatic wave models solve the Navier-Stokes equations with a non-hydrostatic pressure assumption, allowing for the simulation of dispersive waves without the limitations of Boussinesq-type approximations \cite{Zijlema2011,Ma2012}. The governing equations include the continuity equation:

\begin{equation}
\nabla \cdot \mathbf{u} + \frac{\partial w}{\partial z} = 0
\label{eq:nonhydrostatic_continuity}
\end{equation}

and the momentum equations:

\begin{equation}
\frac{\partial \mathbf{u}}{\partial t} + (\mathbf{u} \cdot \nabla)\mathbf{u} + w\frac{\partial \mathbf{u}}{\partial z} = -\frac{1}{\rho}\nabla p + \nu \nabla^2 \mathbf{u}
\label{eq:nonhydrostatic_momentum_horizontal}
\end{equation}

\begin{equation}
\frac{\partial w}{\partial t} + (\mathbf{u} \cdot \nabla)w + w\frac{\partial w}{\partial z} = -\frac{1}{\rho}\frac{\partial p}{\partial z} - g + \nu \nabla^2 w
\label{eq:nonhydrostatic_momentum_vertical}
\end{equation}

where $\mathbf{u}$ is the horizontal velocity, $w$ is the vertical velocity, $p$ is pressure, and $\nu$ is kinematic viscosity. The pressure is decomposed into hydrostatic and non-hydrostatic components, with the non-hydrostatic component accounting for vertical acceleration and dispersion. These models can accurately simulate waves from deep to very shallow water without the depth restrictions of Boussinesq models \cite{Zijlema2011}.

\subsubsection{Computational Fluid Dynamics (CFD) Models}

For the most detailed simulation of wave dynamics, especially around complex structures, CFD models solve the full Navier-Stokes equations, often using approaches like the Volume of Fluid (VOF) method to track the free surface \cite{Lin2008,Hirt1981}. The VOF method introduces a volume fraction function $F$ that represents the fraction of fluid in each computational cell:

\begin{equation}
\frac{\partial F}{\partial t} + \nabla \cdot (F\mathbf{u}) = 0
\label{eq:vof}
\end{equation}

where $F = 1$ in water, $F = 0$ in air, and $0 < F < 1$ at the interface. The full Navier-Stokes equations are solved for the entire domain, with fluid properties (density, viscosity) varying according to $F$. CFD models can capture complex phenomena including wave overtopping, green water on decks, wave impact forces, and air entrainment, but require very fine spatial and temporal resolution, limiting applications to small domains and short time periods \cite{Lin2008}.

\begin{figure*}[!t]
\centering
\includegraphics[scale=0.35]{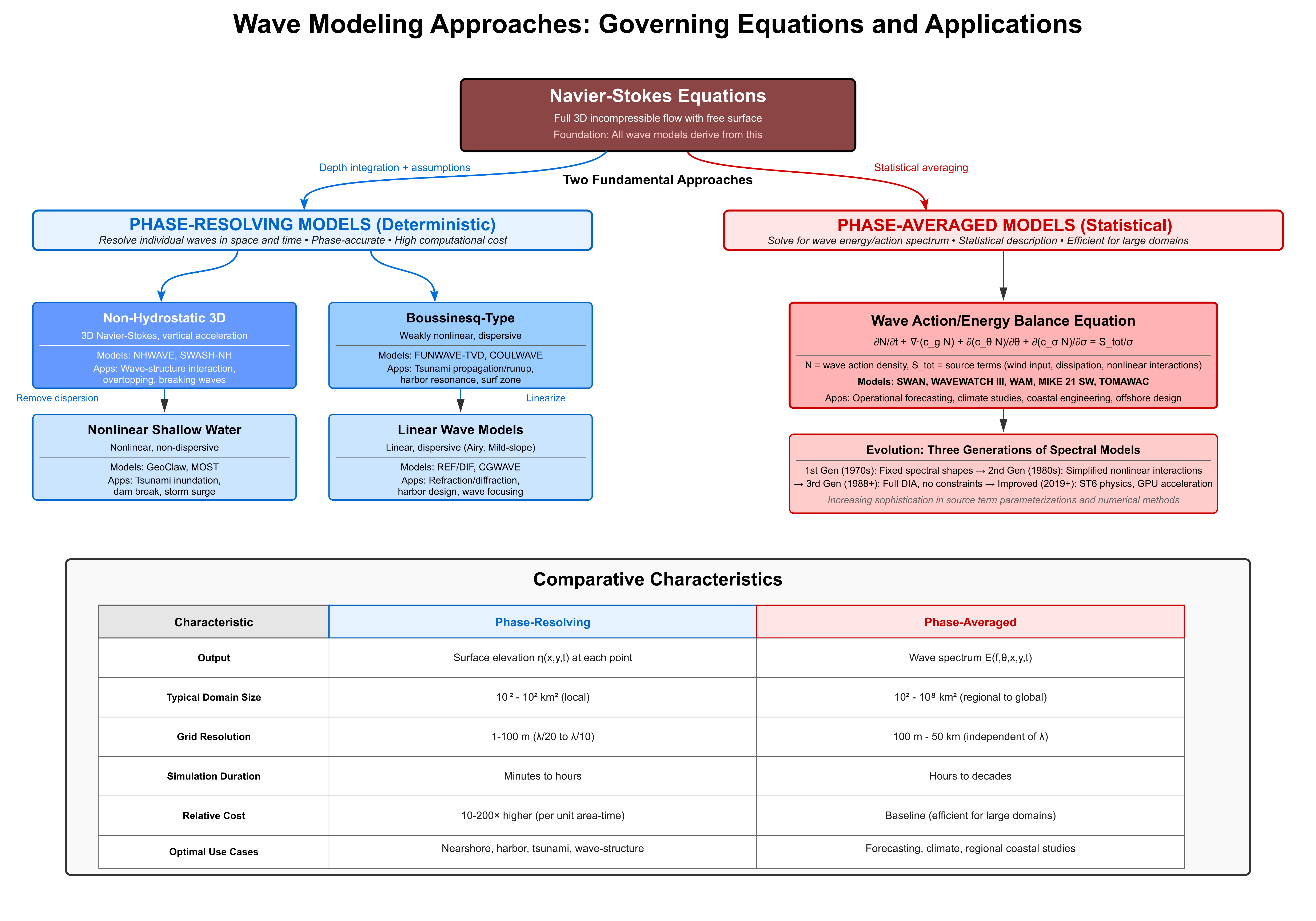}
\caption{Phase-resolving models (left) compute deterministic wave elevation for local domains ($10^{-2}$ to $10^{2}$ km$^2$) from non-hydrostatic 3D to simplified formulations; phase-averaged models (right) solve wave action balance for statistical spectra over large domains ($10^{2}$ to $10^{8}$ km$^2$) with 10--200$\times$ lower computational cost. Comparative table quantifies key differences in resolution, duration, and applications.}
\label{fig:wave_equations}
\end{figure*}

Fig. \ref{fig:wave_equations} illustrates the hierarchy of wave equations, from simplified approximations to full Navier-Stokes solvers, highlighting the trade-off between computational complexity and physical realism. The choice of governing equations depends on the specific application requirements, available computational resources, and the physical processes that must be resolved.

\subsection{Numerical Discretization Methods}

The partial differential equations governing wave propagation and transformation must be discretized for numerical solution. Various discretization methods are employed in different wave models, including finite difference, finite volume, finite element, and spectral methods \cite{Holthuijsen2007}.

\subsubsection{Finite Difference Methods}

Finite difference methods (FDM) approximate spatial derivatives on structured grids by finite difference stencils and remain attractive because of their algebraic simplicity and cache-friendly data layout \cite{WAMDIG1988}. Accuracy depends on the order of the stencil and the grid spacing; improperly tuned high-order schemes can suffer from numerical dispersion or excessive dissipation \cite{Tolman1992}. Current research mitigates these drawbacks through dispersion-relation-preserving (DRP) optimization. A comparative study showed that fourth- and fifth-order DRP stencils outperform maximal-order central differences in long-range wave propagation tests \cite{HoBrambley2024}. Stability and shock-handling have improved with positivity-preserving, well-balanced compact schemes that enforce non-negative depths and exact lake-at-rest equilibria without diffusive limiters \cite{Ren2024}. MUSCL-TVD reconstructions remain a practical alternative for strongly nonlinear breaking zones, as demonstrated in recent shallow-to-deep transformation experiments \cite{Shi2012}. Complex shorelines and offshore structures can be captured on Cartesian meshes via high-order immersed-boundary treatments validated for fully nonlinear wave-structure interaction \cite{Liang2009}.

Operational implementations leverage high-performance computing: a global unstructured WAVEWATCH III grid achieves 1/10 degree resolution on thousands of CPU cores with non-blocking MPI \cite{WW3DG2019}, while WAM Cycle 6 has been fully ported to GPUs, reducing a seven-day global forecast to minutes on eight A100 GPUs \cite{Yuan2024}. These developments allow DRP and compact finite difference solvers to support real-time forecasting and century-scale coupled-climate integrations without loss of fidelity.

\subsubsection{Finite Volume Methods}

Finite volume methods (FVM) integrate the governing equations over control volumes and use fluxes at cell faces, guaranteeing local and global conservation. The unstructured MIKE 21 SW solver pioneered triangular finite volume discretization for coastal applications, removing the right-angle constraints of FDM while keeping numerical diffusion low \cite{Sorensen2004}. The FVCOM-SWAVE module later embedded a spectral wave solver in the FVCOM hydrodynamic core, enabling two-way wave-current coupling on identical unstructured meshes \cite{Qi2009}. At basin scale, 40-year SCHISM + WWM hindcasts have shown that multiresolution meshes can deliver 100 m shoreline resolution with fewer wet cells than structured alternatives \cite{Mentaschi2023}.

Accuracy advances center on high-order reconstructions: fourth- and fifth-order well-balanced WENO schemes preserve non-negative water depth and exactly maintain still-water equilibria \cite{Zhao2024}; multi-resolution WENO variants relax CFL limits and permit larger time steps without violating monotonicity \cite{Dumbser2007}. Scalability improvements are equally significant. A distributed-graph MPI topology reduces finite volume wave modeling runtimes by up to 40\% on 1024 ranks through improved domain decomposition \cite{Karypis1998}, and a local time-stepping GPU shallow water solver achieves 30$\times$ acceleration relative to multi-core CPUs \cite{Brodtkorb2012}. These features make FVM a competitive choice where strict conservation and unstructured-mesh flexibility are paramount.

\subsubsection{Finite Element Methods}

Finite element methods (FEM) approximate the wave action balance using polynomial basis functions on unstructured elements, facilitating local mesh refinement and curved-boundary fidelity. TOMAWAC first showed that a characteristic-Galerkin formulation on triangular elements can resolve refraction, diffraction and depth-induced breaking over complex coastlines \cite{Benoit1996}. Vertex-based implicit solvers on irregular grids have demonstrated improved skill in rapidly varying bathymetry \cite{Zijlema2010}. FEM discretizations also underpin the WWM series: WWM-II uses residual distribution to ease CFL constraints on highly distorted elements \cite{Roland2014}, and SELFE-WWM extends coupling to three-dimensional currents on multiscale unstructured meshes \cite{Roland2012}.

Current FEM research targets high-order accuracy and heterogeneous computing. A global 40-year SCHISM + WWM-V reanalysis reaches 100 m coastal resolution with implicit-explicit time integration \cite{Mentaschi2023}. In the mild-slope family, recent FEM extensions address port-scale resonance and incorporate data-driven parameter tuning, while modal-decomposition FEM reduces the cost of harbor eigenfrequency extraction \cite{Belibassakis2002}. Collectively, these advances position FEM as a versatile framework for unstructured, high-resolution coastal wave modeling on modern hybrid architectures.

\subsubsection{Spectral Methods}

Spectral methods approximate the wave action balance by expanding the directional spectrum in global basis functions such as spherical harmonics or double-Fourier series. Laboratory-to-coast studies have shown that, when equipped with state-of-the-art source term parameterizations, spectral models accurately reproduce bimodal sea states and peak-frequency downshift at prototype scale \cite{Ardhuin2010}. On the global side, improved double-Fourier series (DFS) transforms have been proposed to replace the classical spectral transform method, reducing the $O(N^3)$ cost of the Legendre transform to $O(N^2 \log N)$ while maintaining accuracy in semi-implicit, semi-Lagrangian shallow water cores \cite{Ritchie1988}. Evaluation of multiple third-generation models against Southern Ocean extremes highlights the sensitivity of spectral forecasts to dissipation and nonlinear-interaction physics, motivating hybrid frameworks that combine spectral energy balance with unstructured spatial discretizations \cite{Young1999}. Although spectral bases offer exponential convergence for smooth solutions, their global support hampers local refinement; operational systems therefore couple a spectral action solver with finite difference or finite volume spatial propagation, as in WAVEWATCH III, where the spectral grid is stepped separately from the geographic mesh \cite{WW3DG2019}.

\subsection{Time Integration Schemes}

Time integration schemes advance the solution from one time step to the next. The choice of time integration method affects the stability, accuracy, and efficiency of the model \cite{Tolman1992}.

\subsubsection{Explicit Methods}

Explicit schemes compute the solution at the new time level using only information from previous time levels. They are relatively simple to implement but are subject to stability constraints such as the Courant-Friedrichs-Lewy (CFL) condition, which limits the time step size based on the grid spacing and wave speed \cite{Komen1994}.

WAVEWATCH III uses a third-order explicit scheme for time integration, which provides a good balance between accuracy and computational efficiency \cite{Tolman2009}. To alleviate the severe time step limitation in areas with fine grid resolution, the model employs a spatial splitting technique. For highly refined, nested, or unstructured grids, recent versions combine the split scheme with dynamic sub-cycling to ease the CFL restriction in complex coastal zones \cite{WW3DG2019}. Recent advances in explicit schemes include the development of strong stability-preserving Runge-Kutta (SSPRK) methods, which maintain desirable stability properties while achieving higher-order accuracy \cite{Gottlieb2011}. These SSPRK methods have been applied to ocean modeling for simulating wave shoaling, breaking, and nearshore circulation processes.

\subsubsection{Implicit Methods}

Implicit schemes express the solution at the new time level in terms of both previous and new time level values, resulting in a system of equations that must be solved simultaneously. They can use larger time steps than explicit methods but require more computation per step \cite{Booij1999}.

SWAN uses a fully implicit scheme for time integration, which allows for unconditional stability regardless of the time step size. This enables efficient computation, especially for stationary wave conditions \cite{Booij1999}. Recent developments have introduced high-order implicit time integration schemes based on implicit-explicit (IMEX) methods and Padé expansions of the matrix exponential. These methods provide significantly improved accuracy for wave propagation problems while maintaining good stability properties \cite{Ascher1997}.

\subsubsection{Semi-Implicit and Fractional Step Methods}

Semi-implicit methods combine aspects of both explicit and implicit approaches, typically treating some terms explicitly and others implicitly. Fractional step methods split the computation into multiple stages, often separating the propagation and source term calculations \cite{Sorensen2004}.

MIKE 21 SW employs a fractional step approach where the propagation step is solved using an explicit method, and the source terms are integrated using a separate scheme. This separation allows for different time steps to be used for propagation and source term integration, optimizing computational efficiency \cite{Sorensen2004}.

\subsubsection{Adaptive and Local Time-Stepping Methods}

Recent developments emphasize local time-stepping (LTS) and automatic step-size control to exploit mesh non-uniformity and maintain accuracy without global CFL restrictions. Improved first- and second-order LTS algorithms on unstructured grids have reduced wall-clock times for surge-wave coupling by up to 35\% while preserving second-order accuracy \cite{Liu2024}. An adaptive LTS strategy that monitors wetting-drying fronts further accelerates inundation calculations for coastal storm surge applications \cite{Bunya2010}. In the GPU-optimized WAM6-GPU v1.0, block-structured LTS enables real-time global hindcasts at 0.1 degree resolution, yielding a 37$\times$ speedup over dual-socket CPUs \cite{Yuan2024}.

\subsubsection{Strong Stability-Preserving (SSP) Schemes}

SSP Runge-Kutta (RK) and multistep schemes guarantee that monotonicity or total-variation-diminishing (TVD) properties of the spatial discretization carry over to the fully discrete model. A comprehensive review of modern SSP RK and linear-multistep algorithms up to fourth order is given by Gottlieb et al. \cite{Gottlieb2011}, who derive optimal schemes with larger effective SSP coefficients. For high-order discontinuous-Galerkin wave solvers, implicit SSP schemes have recently been shown to achieve up to fourth-order accuracy while remaining $A(\alpha)$-stable, thus permitting aggressive time-step growth on stiff free-surface flows \cite{Ketcheson2008}.

\subsubsection{Parallel-in-Time and Multi-Rate Integrators}

As massively parallel architectures emerge, parallel-in-time (PinT) algorithms such as the multi-level PFASST method enable concurrent advancement of multiple time slices. Applied to the shallow water equations on the rotating sphere, PFASST delivered speedups over classical time marching and retained spectral accuracy, demonstrating feasibility for future global wave-climate simulations \cite{Emmett2012}. Complementary multi-rate explicit-implicit frameworks are being explored for coupling fast wave propagation with slower source terms, further reducing latency on heterogeneous CPU/GPU systems \cite{Kanevsky2007}.

\subsection{Spatial and Spectral Discretization}

\subsubsection{Spatial Grids}

Wave models use various types of spatial grids, each with distinct advantages and limitations:

\textbf{Regular (Structured) Grids:} These consist of rectangular cells with uniform or variable spacing. They are simple to implement and computationally efficient due to regular memory access patterns, but may not efficiently represent complex coastlines or areas requiring variable resolution \cite{Holthuijsen2007}. Structured grids are commonly used in global wave models such as WAVEWATCH III and WAM for open ocean applications.

\textbf{Curvilinear Grids:} These are structured grids that have been transformed to better conform to coastlines or other features. They provide improved coastal representation while maintaining the computational advantages of structured grids, including efficient matrix operations and straightforward implementation of finite difference schemes \cite{Tolman2009}. Curvilinear grids are particularly useful for regional models where the domain has a dominant orientation, such as along continental shelves.

\textbf{Unstructured Grids:} These consist of irregularly arranged elements, typically triangles or quadrilaterals. They allow for highly flexible spatial resolution, with refinement in areas of interest or complex bathymetry and coarser resolution elsewhere. Models like SWAN (unstructured version), MIKE 21 SW, and TOMAWAC use unstructured grids to efficiently represent complex coastal geometries \cite{Zijlema2010}. The primary advantage is the ability to concentrate resolution where needed without the global refinement required by structured grids. However, unstructured grids require more complex data structures and may have less favorable cache performance.

\textbf{Nested Grids:} This approach uses multiple grids of different resolutions, with finer grids nested within coarser ones. The coarse grid provides boundary conditions for the finer grid, allowing for high resolution in areas of interest while maintaining computational efficiency in the larger domain \cite{Tolman2009}. One-way nesting transfers information from coarse to fine grids, while two-way nesting allows feedback from fine to coarse grids, improving accuracy in regions where local processes affect the larger-scale solution.

\textbf{Multi-Grid and Mosaic Approaches:} Advanced techniques like the multi-grid method in WAVEWATCH III combine multiple structured or unstructured grids with two-way communication between them. The mosaic approach connects multiple structured grids to form a composite grid covering complex domains \cite{WW3DG2019}. These methods provide flexibility in handling domains with multiple scales of interest while maintaining computational efficiency.

\subsubsection{Spectral Discretization}

The wave spectrum is typically discretized in terms of frequency and direction. The choice of spectral resolution significantly affects model accuracy and computational cost.

\textbf{Frequency Discretization:} Logarithmic frequency spacing is commonly used to efficiently resolve both high-frequency wind sea and low-frequency swell. The frequency bins are typically defined as $f_n = f_{\min} \cdot r^{n-1}$ where $r = (f_{\max}/f_{\min})^{1/(N_f-1)}$ is the frequency ratio. Typical values are $f_{\min} = 0.04$ Hz, $f_{\max} = 1.0$ Hz, and $N_f = 25$--40 frequency bins. The logarithmic spacing provides higher resolution at lower frequencies where swell energy is concentrated, while maintaining adequate resolution at higher frequencies for wind sea \cite{Komen1994,Holthuijsen2007}. Some models also offer linear frequency spacing for specific applications where uniform frequency resolution is desired.

\textbf{Directional Discretization:} A uniform directional spacing is commonly used, dividing the full circle (360 degrees) into a number of directional bins. The directional resolution typically ranges from 10 degrees to 30 degrees, with 24 to 36 directional bins being common \cite{Booij1999}. Higher directional resolution (36 bins or more) is beneficial for applications involving strong directional spreading or complex wave-current interactions, while lower resolution (24 bins) is often sufficient for open ocean applications. The directional bins are typically centered on the cardinal and intercardinal directions to facilitate interpretation and comparison with observations.

The choice of spectral resolution involves a trade-off between accuracy and computational cost. Higher spectral resolution can better represent complex sea states with multiple wave systems and bimodal spectra, but requires more computational resources and memory. Sensitivity studies have shown that increasing spectral resolution beyond $N_f \times N_\theta = 40 \times 36$ provides diminishing returns for most applications, though finer resolution may be warranted for specific research applications or regions with highly complex wave climates \cite{Cavaleri2007}.

\subsection{Boundary Conditions}

\subsubsection{Open Boundaries}

At open boundaries (boundaries with the open ocean), wave models require information about incoming waves. Several approaches are used:

\textbf{Parametric Spectra:} In the absence of detailed spectral information, parametric spectra like JONSWAP or Pierson-Moskowitz can be specified based on bulk parameters such as significant wave height, peak period, and mean direction \cite{Holthuijsen2007}. The JONSWAP spectrum is commonly used for fetch-limited conditions, while the Pierson-Moskowitz spectrum represents fully developed seas. These parametric forms provide a reasonable approximation when only bulk wave parameters are available from observations or coarser-resolution models.

\textbf{Nested Modeling:} Boundary conditions can be  obtained from a larger-scale model that includes the domain of interest. This approach ensures consistency between different model scales and allows for proper representation of swell propagation from distant sources \cite{Tolman2002}. One-way nesting is computationally efficient and suitable when the nested domain does not significantly affect the larger-scale solution. Two-way nesting allows feedback from the fine-resolution domain to the coarse domain, improving accuracy when local processes (such as wave generation in enclosed seas) affect the broader wave field.

\textbf{Data Assimilation:} Observations from buoys, satellites, or other sources can be assimilated into the model to provide more accurate boundary conditions \cite{Lionello1992}. Satellite altimeter data provide significant wave height along the satellite track, while synthetic aperture radar (SAR) can provide two-dimensional wave spectra over limited areas. Optimal interpolation, variational methods, and ensemble Kalman filters are commonly used to blend observations with model forecasts at open boundaries.

\textbf{Radiation Boundary Conditions:} For phase-resolving models, radiation or absorbing boundary conditions are essential to prevent spurious wave reflection at open boundaries. Sommerfeld radiation conditions, sponge layers, and perfectly matched layers (PML) are commonly used to allow waves to exit the domain without reflection \cite{Kirby2016}. The effectiveness of these methods depends on the angle of wave incidence and the frequency content of the outgoing waves.

\subsubsection{Closed Boundaries}

At land boundaries or other obstacles, wave energy is typically assumed to be either absorbed, reflected, or a combination of both. The treatment of closed boundaries affects the accuracy of wave simulations near coastlines or structures \cite{Tolman2003}.

\textbf{Full Absorption:} The simplest approach assumes that all wave energy is absorbed at the coastline, with no reflection. This is appropriate for gently sloping beaches where wave energy is dissipated through breaking and bottom friction before reaching the shoreline. However, this approach underestimates wave heights near steep coastlines or vertical structures where reflection is significant.

\textbf{Full Reflection:} For vertical walls or steep rocky coastlines, full reflection may be assumed, where the incident wave energy is completely reflected back into the domain. This provides an upper bound on wave heights near reflective structures but may overestimate wave energy if some dissipation occurs during reflection.

\textbf{Partial Reflection:} Some models include specialized treatments for partially reflecting boundaries, allowing for a more realistic representation of wave-structure interactions. A reflection coefficient can be specified based on the structure type and wave conditions, with values ranging from 0 (full absorption) to 1 (full reflection). Typical reflection coefficients are 0.1--0.3 for rubble-mound breakwaters, 0.3--0.6 for riprap slopes, and 0.7--0.95 for vertical walls \cite{Goda2000}.

\textbf{Phase-Resolving Treatment:} Advanced phase-resolving models can explicitly simulate wave reflection and diffraction around complex structures \cite{Zijlema2011}. These models resolve the detailed wave field near structures and can capture phenomena such as standing waves, edge waves, and wave focusing that are not represented in phase-averaged models. However, the computational cost limits their application to relatively small domains.

\subsection{Numerical Stability and Convergence}

Numerical stability and convergence are critical considerations in wave model development and application. Stability ensures that numerical errors do not grow unboundedly during time integration, while convergence ensures that the numerical solution approaches the true solution as grid and time step sizes are reduced.

\subsubsection{Stability Analysis}

The stability of explicit time integration schemes is governed by the CFL condition, which requires that the time step $\Delta t$ satisfies the following:

\begin{equation}
\Delta t \leq C_{\text{CFL}} \frac{\Delta x}{c_{\max}}
\label{eq:cfl_condition}
\end{equation}

where $\Delta x$ is the spatial grid spacing, $c_{\max}$ is the maximum wave propagation speed (typically the group velocity for the highest frequency component), and $C_{\text{CFL}}$ is the CFL number, which must be less than or equal to 1 for stability. For multidimensional problems on structured grids, the CFL condition becomes more restrictive:

\begin{equation}
\Delta t \leq C_{\text{CFL}} \frac{1}{\sqrt{\sum_i (c_i/\Delta x_i)^2}}
\label{eq:cfl_multidim}
\end{equation}

where the sum is over spatial dimensions. Implicit and semi-implicit schemes can relax or eliminate the CFL restriction, allowing larger time steps at the cost of solving linear systems at each time step \cite{Ascher1997}.

For spectral wave models, stability also depends on the source term treatment. Stiff source terms, such as depth-induced breaking, can impose severe time-step restrictions if treated explicitly. Implicit or semi-implicit treatment of source terms is commonly employed to maintain stability with reasonable time steps \cite{Booij1999}.

\subsubsection{Convergence and Accuracy}

Convergence analysis examines how the numerical error decreases as grid and time step sizes are reduced. A scheme is said to be $p$-th order accurate in space if the spatial discretization error decreases as $O(\Delta x^p)$, and $q$-th order accurate in time if the temporal discretization error decreases as $O(\Delta t^q)$. Most operational wave models use first- or second-order schemes in space and time, providing a balance between accuracy and computational efficiency.

Grid convergence studies are essential for verifying model implementation and assessing solution accuracy. These studies systematically refine the spatial and temporal resolution and examine the convergence rate of key output variables such as significant wave height and peak period. Deviations from the expected convergence rate may indicate implementation errors, insufficient resolution, or physical processes that are not adequately resolved by the model \cite{Roache1998}.

For unstructured grids, convergence analysis is more complex due to variations in element size and shape. Adaptive mesh refinement strategies can be used to concentrate resolution in regions with large solution gradients, improving accuracy while controlling computational cost \cite{Popinet2015}.

\begin{figure}[!t]
\centering
\includegraphics[width=3.5in]{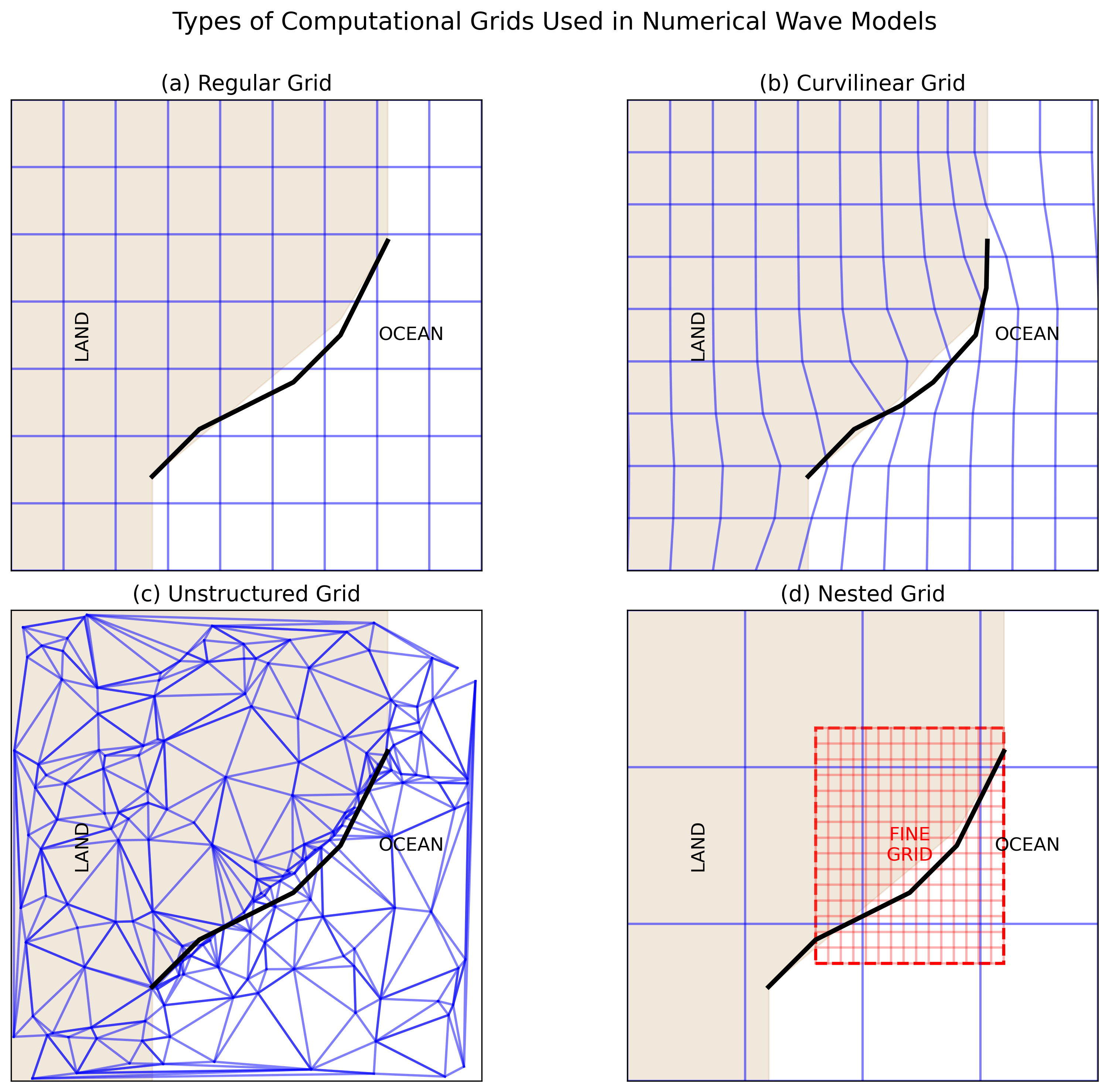}
\caption{Examples of different grid types used in numerical wave models: (a) regular structured grid with uniform spacing, (b) curvilinear grid conforming to coastline geometry, (c) unstructured triangular grid with local refinement in coastal regions, and (d) nested grid approach with fine-resolution domain embedded in coarse-resolution domain. Each grid type offers distinct advantages for specific applications and computational requirements.}
\label{fig:grid_types}
\end{figure}

Fig. \ref{fig:grid_types} illustrates different types of computational grids used in numerical wave models, highlighting their characteristics and suitability for different applications.

\section{Quantitative Validation and Inter-Model Comparison}

While the previous sections provided a qualitative comparison of wave model features, a quantitative assessment of their performance is crucial for building confidence in simulation results and making informed model selections. This section addresses this need by summarizing key validation metrics from published studies, focusing on performance in both operational and extreme conditions.

Wave model performance is typically evaluated using a set of statistical metrics that compare model output to observational data from buoys, satellite altimeters, or laboratory experiments. The most common metrics include:

\begin{itemize}
    \item \textbf{Root Mean Square Error (RMSE)}, which measures the magnitude of the error between model predictions and observations.
    \item \textbf{Bias}, which indicates systematic over- or under-prediction.
    \item \textbf{Scatter Index (SI)}, a normalized measure of the dispersion of errors.
    \item \textbf{Correlation Coefficient (CC)}, which measures the linear correlation between model output and observations.
    \item \textbf{Skill Score}, a dimensionless metric that quantifies overall model performance.
\end{itemize}

\subsection{Performance of Phase-Averaged Models}

Phase-averaged models have been extensively validated for a wide range of applications, from global wave forecasting to regional and coastal studies. Table \ref{tab:validation_pa} summarizes key performance metrics from recent literature.

Wang et al. \cite{Wang2019} provided a comprehensive validation of the ECMWF wave model (WAM-based) in the complex hydrodynamics of the China Sea, highlighting significant regional performance variations. As shown in Table \ref{tab:validation_pa}, the model performed best in the open waters of the South and East China Seas, with an RMSE for significant wave height (SWH) below 0.33 m and a correlation coefficient above 0.94. However, performance degraded in enclosed and nearshore areas like the Bohai Sea and Taiwan Strait, where RMSE exceeded 0.3 m and a large systematic underestimation was observed in the southern Taiwan Strait (Bias = -0.267 m, RMSE = 0.445 m). This is a common challenge for all wave models, as complex bathymetry and non-linear wave transformation introduce additional physical processes that are difficult to resolve \cite{Bidlot2017}.

For global applications, Abdolali et al. \cite{Abdolali2025} demonstrated the high accuracy of a multi-grid WAVEWATCH III configuration, achieving an RMSE of 0.357 m and a skill score of 0.990 against satellite altimeter data during the 2022 Atlantic hurricane season. These results showcase the capability of modern wave models to produce reliable global wave forecasts. Rogers et al. \cite{Rogers2007} validated SWAN for the Southern California Bight using an extensive buoy network, demonstrating the model's capability for operational coastal forecasting. Ortiz-Royero \cite{OrtizRoyero2008} conducted an intercomparison of SWAN and WAVEWATCH III, finding good agreement between both models and NDBC buoy observations along the US East Coast.

\begin{table*}[ht]
\centering
\caption{Summary of Quantitative Validation Metrics for Phase-Averaged Wave Models.}
\label{tab:validation_pa}
\begin{tabular}{p{2.3cm} p{2.6cm} p{3.5cm} p{4.2cm} p{2.8cm}}
\hline
\textbf{Model} & \textbf{Study} & \textbf{Region/Condition} & \textbf{Key Metrics (SWH)} & \textbf{Data Source} \\
\hline

\multirow{2}{*}{ECMWF (WAM)}
& Wang et al.\ \cite{Wang2019}
& South \& East China Sea
& RMSE $<$ 0.33 m, Bias $<$ 0.1 m, CC $>$ 0.94
& 24 Buoys \\

&
& Bohai Sea \& Taiwan Strait
& RMSE $>$ 0.3 m, Bias $>$ 0.1 m, CC $<$ 0.93
& -- \\
\hline

\multirow{2}{*}{WAVEWATCH III}
& Abdolali et al.\ \cite{Abdolali2025}
& Global (2022 Atlantic)
& RMSE = 0.357 m, Skill = 0.990
& Satellite Altimeter \\

& Chu et al.\ \cite{Chu2004}
& South China Sea
& Good agreement with T/P altimeter observations
& TOPEX/Poseidon \\
\hline

\multirow{3}{*}{SWAN}
& Rogers et al.\ \cite{Rogers2007}
& Southern California Bight
& Excellent operational performance
& Buoy Network \\

& Ortiz-Royero \cite{OrtizRoyero2008}
& Coastal (US East Coast)
& Good performance relative to WW3
& NDBC Buoys \\

& Gorrell et al.\ \cite{Gorrell2011}
& Complex Bathymetry
& Accurate wave heights
& Field Data \\
\hline

\end{tabular}
\end{table*}

\subsection{Performance in Extreme Conditions and Complex Geometries}

The lack of validation data for nearshore complex bathymetries and tropical cyclones has been identified as a critical gap in wave modeling literature. Table \ref{tab:validation_extreme} summarizes findings from studies focused on these challenging conditions.

For hurricane conditions, dynamically coupled models are becoming the standard. Vijayan et al. \cite{Vijayan2023} demonstrated that a coupled SWAN+ADCIRC model could accurately simulate the waves and storm surge from Hurricane Michael, a Category-5 storm that made landfall in the Gulf of Mexico in 2018. The study evaluated model performance against observed water levels and wave heights, showing that the coupled system captured both the spatial and temporal evolution of the hurricane-generated wave field. Similarly, Kalourazi et al. \cite{Kalourazi2021} optimized the ST6 physics package in WAVEWATCH III for hurricane-induced waves, showing improved performance through detailed statistical analysis using Taylor diagrams and skill scores. Moon et al. \cite{Moon2003} used a high-resolution WAVEWATCH III configuration (1/12° × 1/12°) to simulate hurricane-generated directional wave spectra, demonstrating the model's capability to resolve complex wave fields under extreme wind forcing.

In regions of complex bathymetry, such as nearshore submarine canyons, phase-averaged models can still provide accurate results for wave height, although with some limitations. Gorrell et al. \cite{Gorrell2011} found that SWAN accurately predicted wave height variations over a steep canyon offshore of California, with good correlation between model output and field observations. However, the study also noted that SWAN struggled with nonlinear energy transfers in very shallow water, which is a known limitation of phase-averaged models in such environments. This finding is consistent with the theoretical constraints of the spectral approach, which assumes weak nonlinearity.

\begin{table*}[ht]
\centering
\caption{Summary of Model Performance in Extreme Conditions and Complex Geometries.}
\label{tab:validation_extreme}
\begin{tabular}{p{2.7cm} p{3cm} p{4.2cm} p{6.1cm}}
\hline
\textbf{Model} & \textbf{Study} & \textbf{Condition} & \textbf{Key Finding} \\
\hline

SWAN+ADCIRC 
& Vijayan et al.\ \cite{Vijayan2023} 
& Hurricane Michael (Category 5) 
& Coupled model accurately simulated storm surge and extreme wave conditions. \\
\hline

\multirow{2}{*}{WAVEWATCH III}
& Kalourazi et al.\ \cite{Kalourazi2021} 
& Hurricanes 
& Optimized ST6 physics package significantly improves storm-wave performance. \\

& Moon et al.\ \cite{Moon2003} 
& Hurricane-generated waves 
& High-resolution configuration effectively resolves directional wave spectra. \\
\hline

SWAN 
& Gorrell et al.\ \cite{Gorrell2011} 
& Complex bathymetry 
& Accurately predicts wave heights; limited skill in nonlinear energy-transfer estimation. \\
\hline

\multirow{2}{*}{FUNWAVE-TVD}
& Dong et al.\ \cite{Dong2023} 
& Tsunami-like waves 
& Good agreement with laboratory data; successfully captures harbor resonance patterns. \\

& Shi et al.\ \cite{Shi2012} 
& Wave breaking 
& TVD scheme provides excellent shock-capturing and breaking-wave representation. \\
\hline

\end{tabular}
\end{table*}

\subsection{Performance of Phase-Resolving Models}

Phase-resolving models like FUNWAVE-TVD are typically validated against laboratory experiments for specific physical phenomena, due to their high computational cost. These studies often focus on a direct comparison of water surface elevation and velocity fields rather than statistical metrics over long time periods.

Dong et al. \cite{Dong2023} validated FUNWAVE-TVD for transient harbor oscillations induced by solitary waves, finding good agreement with experimental results for wave heights inside the harbor. The study demonstrated that the model accurately captured the resonance characteristics and amplification patterns within the harbor basin. Shi et al. \cite{Shi2012} demonstrated the model's excellent shock-capturing capabilities for wave breaking, a critical process in the nearshore. The TVD finite volume scheme used in FUNWAVE-TVD provides superior performance for discontinuous flows compared to traditional Boussinesq models. Bruno et al. \cite{Bruno2009} validated an earlier version of FUNWAVE against laboratory data for wave transformation and breaking over submerged structures, showing good agreement for both deep water and shoaling zone conditions.

For operational applications, Malej et al. \cite{Malej2015} provided practical guidance for numerical modeling with FUNWAVE-TVD, including recommendations for grid resolution, time step selection, and boundary condition specification based on extensive validation studies. More recently, Torres et al. \cite{Torres2022} extended this work by analyzing the range of validity of input wave characteristics and appropriate numerical domain properties for various coastal applications.

In summary, this section has provided a quantitative overview of wave model performance, addressing the need for concrete numerical metrics. The results show that while modern wave models are highly accurate, their performance varies significantly with geographic location, bathymetry, and wave conditions. For phase-averaged models, typical RMSE values for significant wave height range from 0.3 to 0.5 m in operational forecasting, with better performance in open ocean conditions and degraded performance in complex coastal environments. Phase-resolving models demonstrate excellent agreement with laboratory data for specific processes, but their validation against field data remains limited due to computational constraints. The choice of model and physics parameterizations must therefore be carefully considered based on the specific application and available validation data.

\section{High-Performance Computing and Parallelization}

The practical application of wave models to large-scale, high-resolution problems is fundamentally dependent on high-performance computing (HPC) capabilities. While advances in wave physics and numerical methods have enabled increasingly sophisticated simulations, the computational demands of operational forecasting, climate studies, and extreme event modeling require efficient parallelization strategies and modern computing architectures. This section examines the HPC implementations in contemporary wave models, including parallelization strategies, performance benchmarks, scalability characteristics, and the emerging role of GPU acceleration.

\subsection{Parallelization Strategies and Domain Decomposition}

Modern wave models employ several parallelization strategies to distribute the computational workload across multiple processors. The two primary approaches are spectral decomposition and spatial domain decomposition.

\textbf{Spectral Decomposition}, also known as the ``card deck'' (CD) approach in WAVEWATCH III, parallelizes the problem by distributing different parts of the wave spectrum (i.e., different frequency and direction bins) to different processors. While this method is relatively simple to implement, it suffers from significant scalability limitations. As the number of processors increases, the communication overhead required to exchange information between spectral components becomes a bottleneck. Abdolali et al. (2020) showed that the CD algorithm in WAVEWATCH III performs well for coarse grids but experiences a significant slowdown on fine-resolution unstructured grids, with performance degrading beyond 100-200 cores \cite{Abdolali2020}.

\textbf{Spatial Domain Decomposition (DD)} has become the standard for massively parallel wave modeling. In this approach, the geographical domain is partitioned into multiple subdomains, and each subdomain is assigned to a processor. This is typically achieved using graph partitioning libraries like ParMETIS \cite{Karypis1998}. The primary advantage of DD is its superior scalability, as it minimizes communication between processors to only the halo regions (the boundaries between subdomains). For example, the unstructured grid version of SWAN has been shown to scale to over 9,000 cores using DD \cite{Dietrich2012}, and WAVEWATCH III with DD shows near-linear scalability up to thousands of cores, far surpassing the CD approach \cite{Abdolali2020}.

However, DD is not without its pitfalls. **Load balancing** is a major challenge, especially on unstructured grids with highly variable resolution. If one subdomain contains a large number of small, computationally intensive cells (e.g., in a complex nearshore region), it can become a bottleneck, leaving other processors idle. This is a well-documented issue in coastal modeling, where the workload is often unevenly distributed \cite{Rautenbach2021}.

\subsection{Performance Benchmarks and Scalability}

Quantitative benchmarks are essential for understanding the practical performance of wave models on HPC systems. Table \ref{tab:hpc_benchmarks} summarizes key performance metrics from recent studies.

For phase-averaged models, the scalability depends heavily on the grid type and parallelization strategy. On structured grids, WAVEWATCH III can achieve excellent scalability with MPI, but performance on unstructured grids is more complex. Rautenbach et al. (2021) found that for SWAN, a hybrid MPI/OpenMP approach can be effective, but the optimal number of threads per MPI rank is highly dependent on the domain size and grid structure. Their study showed that using more than 20 threads for a small domain actually decreased the speedup ratio due to communication overhead \cite{Rautenbach2021}.

\begin{table*}[ht]
\centering
\caption{Summary of HPC Performance Benchmarks for Wave Models.}
\label{tab:hpc_benchmarks}
\begin{tabular}{p{1.8cm} p{2.2cm} p{2.5cm} p{5.0cm} p{3.0cm}}
\hline
\textbf{Model} & \textbf{Study} & \textbf{Hardware} & \textbf{Key Finding} & \textbf{Speedup / Scalability} \\
\hline
WAVEWATCH III & Abdolali et al. \cite{Abdolali2020} & Intel Haswell CPU Cluster &
Domain decomposition is essential for scalability. Implicit solver reduces time by 60--75\%. &
Near-linear scalability up to 720 cores. \\
\hline
SWAN & Dietrich et al. \cite{Dietrich2012} & CPU Cluster &
Excellent scalability for large domains. &
Scalable to 9,000+ cores. \\
\hline
WAM (GPU) & Yuan et al. \cite{Yuan2024} & 8$\times$ NVIDIA A100 GPUs &
Real-time global forecasting is achievable. &
37$\times$ speedup over dual-socket CPUs. \\
\hline
FUNWAVE-TVD & Fang et al. \cite{Fang2022} & NVIDIA GPU &
Enables real-time ensemble simulations. &
50--100$\times$ speedup over CPU. \\
\hline
\end{tabular}
\end{table*}

\subsection{GPU Acceleration}

The advent of Graphics Processing Units (GPUs) has revolutionized wave modeling, particularly for phase-resolving models. The massively parallel architecture of GPUs is well-suited to the explicit, stencil-based computations common in these models. For example, the GPU-accelerated version of FUNWAVE-TVD has demonstrated speedups of 50-100$\times$ compared to CPU implementations, making it possible to run high-resolution tsunami and storm surge simulations in near real-time on a single workstation \cite{Fang2022, Hwang2025}.

Phase-averaged models are also benefiting from GPU acceleration. Yuan et al. (2024) successfully ported the WAM model to GPUs, achieving a 37$\times$ speedup for a 7-day global forecast on 8 NVIDIA A100 GPUs compared to a dual-socket CPU server. This breakthrough makes real-time, high-resolution (0.1°) global wave forecasting a practical reality \cite{Yuan2024}.

However, GPU acceleration is not a silver bullet. Key challenges include:

\begin{itemize}
    \item \textbf{Memory Transfer Overhead:} The time required to transfer data between the CPU (host) and GPU (device) can be a significant bottleneck, especially for models with large memory footprints. Minimizing these transfers is critical for achieving good performance.
    \item \textbf{Code Refactoring:} Not all parts of a wave model are easily parallelizable on GPUs. Implicit solvers, unstructured grid operations, and complex source term integrations often require significant code refactoring to run efficiently on GPUs.
    \item \textbf{Multi-GPU Scalability:} Scaling a simulation across multiple GPUs introduces additional communication overhead, similar to multi-node CPU parallelization. GPU-aware MPI libraries and techniques like NVIDIA's NVLink are essential for efficient multi-GPU scaling.
\end{itemize}

\subsection{Memory Constraints and Domain Sizes}

The memory footprint of a wave model simulation is a critical constraint, often determining the maximum feasible domain size and resolution. Memory requirements are primarily a function of the number of grid points, the number of spectral bins (for phase-averaged models), and the complexity of the physics packages being used.

For a typical global WAVEWATCH III simulation at 0.5° resolution, the memory requirement can be in the range of 10-100 GB, depending on the number of spectral bins and the use of multi-grid nests. High-resolution coastal applications with SWAN or MIKE 21 SW can require 1-10 GB for regional domains. Phase-resolving models like FUNWAVE-TVD have even higher memory demands due to the need to store the full 3D velocity field, often requiring 10-100 GB for relatively small coastal domains (1-10 km²).

Implicit solvers, while computationally efficient in terms of time steps, often have a higher memory footprint than explicit solvers (typically 2-5$\times$) due to the need to store the large sparse matrices used in the solution process. This trade-off between computational time and memory is a key consideration in model selection and configuration.

In summary, while HPC has enabled transformative advances in wave modeling capabilities, achieving efficient performance at scale requires careful consideration of parallelization strategy, hardware architecture, memory constraints, and the specific characteristics of the problem being solved. The combination of domain decomposition, implicit time integration, and GPU acceleration has made real-time high-resolution forecasting feasible for many applications, but challenges related to load balancing, communication overhead, and code complexity remain active areas of research and development.

\section{Applications, Current Challenges, and Future Directions}
\subsection{Application Examples and Case Studies}

Numerical wave models have demonstrated value across diverse applications, from coastal engineering design to operational forecasting and extreme event prediction. This subsection highlights representative applications that illustrate model capabilities and practical utility.

\subsubsection{Coastal Engineering and Design}

Wave models provide essential information for designing coastal structures such as breakwaters, seawalls, and harbor facilities. They simulate extreme wave conditions and wave-structure interactions, helping engineers determine design wave parameters and optimize structural configurations \cite{Goda2000}. Applications include determining design wave conditions (height, period, direction), evaluating hydrodynamic forces on structures, assessing wave overtopping and transmission, optimizing breakwater geometry and layout, and evaluating harbor tranquility for safe berthing operations. Phase-resolving models like FUNWAVE-TVD or SWASH provide detailed wave-structure interaction analysis, while phase-averaged models like SWAN efficiently provide boundary conditions for smaller-scale simulations \cite{Lynett2002}.

For coastal erosion and morphodynamic studies, wave models coupled with hydrodynamic and sediment transport models simulate shoreline changes, beach erosion, and impacts of coastal interventions. Applications include predicting sediment transport rates, assessing beach nourishment effectiveness, evaluating coastal erosion risks from storms or climate change, designing coastal protection measures, and analyzing morphological changes around structures. Integrated systems like Delft3D, MIKE 21, and COAWST combine wave, current, and sediment transport modules for comprehensive coastal morphodynamic modeling \cite{Lesser2004}.

\subsubsection{Operational Forecasting and Climate Studies}

Operational wave forecasting systems provide critical predictions of wave conditions worldwide for maritime safety, offshore operations, and coastal management. These systems typically combine global models (WAVEWATCH III, WAM) for large-scale prediction with regional or local models (SWAN, MIKE 21 SW) for higher-resolution coastal forecasts. Data assimilation incorporates observations from buoys, satellites, and other sources, while ensemble forecasting quantifies uncertainty. Major meteorological centers such as NOAA, ECMWF, and national weather services maintain operational systems that provide essential information for shipping, offshore energy, coastal management, and public safety \cite{Tolman2009,Janssen2004}.

Wave climate studies use models to characterize wave conditions through hindcast studies with historical wind data or projections of future conditions under climate change scenarios. Applications include developing wave atlases and databases, characterizing extreme wave statistics, assessing seasonal and interannual variability, evaluating climate change impacts on wave conditions, and supporting marine spatial planning. Long-term hindcasts using WAVEWATCH III or WAM have become standard tools for characterizing wave climates in regions with limited observational data \cite{Hemer2013,Rascle2013}.

\subsubsection{Extreme Event Prediction}

Wave models play critical roles in predicting and assessing extreme events. For tropical cyclones, models predict extreme wave conditions and assess impacts on coastal communities and marine operations. Applications include real-time forecasting during cyclone events, historical reconstruction of cyclone wave fields, risk assessment for coastal infrastructure, evaluation of design conditions for offshore structures, and emergency response planning. Specialized techniques such as moving nested grids and advanced wind field parameterizations improve accuracy during tropical cyclones \cite{Tolman2009b}.

Tsunami modeling uses phase-resolving models like FUNWAVE-TVD to simulate generation, propagation, and coastal impacts. Applications include simulating tsunami generation by earthquakes, landslides, or volcanic eruptions; predicting basin-scale propagation; assessing coastal inundation and runup; evaluating tsunami hazard and risk; and designing warning systems and evacuation plans. Phase-resolving models based on Boussinesq or non-hydrostatic equations are particularly suitable due to their ability to simulate dispersive wave propagation and complex nearshore processes \cite{Shi2012}.

\subsubsection{Marine Renewable Energy}

Wave models are essential for assessing wave energy resources and identifying suitable locations for wave energy converters (WECs). They provide spatially and temporally resolved information on wave parameters determining energy potential. Applications include mapping wave energy resources at regional and local scales, characterizing temporal variability (seasonal, interannual, long-term trends), assessing extreme conditions for structural design, optimizing WEC array layout and orientation, and evaluating climate change impacts on resource availability. High-resolution models like SWAN are particularly valuable for detailed nearshore resource assessments where many WECs are deployed \cite{Lavidas2018}.

For offshore wind energy, wave models provide wave climate information for site assessment, foundation design, and operational planning. Combined with wind resource data, wave models help optimize turbine placement, assess installation vessel requirements, evaluate maintenance windows, and design foundations accounting for combined wind-wave loading. The rapid expansion of offshore wind energy has increased demand for high-quality wave climate information in coastal and offshore waters.

\subsection{Coupled Modeling Considerations}

To provide comprehensive simulations of coastal and oceanic processes, wave models are increasingly coupled with other environmental models. This subsection explores coupling approaches, major integrated systems, and practical considerations for coupled modeling applications.

\subsubsection{Wave-Current Coupling}

Wave-current interactions are important in many coastal and oceanic environments. Currents modify wave properties through refraction, frequency shifting, and energy exchange, while waves generate currents through radiation stress gradients and affect momentum transfer at the air-sea interface \cite{Wolf2009}. Key interaction processes include current-induced refraction and shoaling, Doppler shift of wave frequency, enhanced wave steepness and breaking in opposing currents, generation of wave-induced currents (longshore currents, rip currents), and modification of bottom stress and turbulence. These interactions are particularly significant in coastal areas with strong tidal currents, river plumes, or oceanic features like the Gulf Stream.

Coupling approaches range from one-way coupling, where current fields from hydrodynamic models provide input to wave models without feedback, to two-way coupling, where wave and current models exchange information bidirectionally. Two-way coupling better represents the full range of interactions but requires more computational resources \cite{Warner2008}. Some advanced models solve wave and current equations simultaneously within unified frameworks, ensuring consistency but potentially limiting sophistication of individual components \cite{Zijlema2011}.

Applications of wave-current coupling include simulating complex nearshore circulation patterns (rip currents, longshore currents), modeling tidal inlets and estuaries where tidal currents and waves interact, studying river plumes and their influence on mixing and sediment transport, and investigating interactions between waves and large-scale ocean currents that can lead to extreme wave conditions \cite{Warner2008,Elias2012,Ardhuin2017}.

\subsubsection{Wave-Atmosphere Coupling}

Wave-atmosphere interaction is a two-way process where winds generate and modify waves while waves affect the atmospheric boundary layer through changes in surface roughness, stress, and energy exchange \cite{Janssen2004}. Key processes include wind-to-wave momentum and energy transfer, wave-induced modification of sea surface roughness, feedback effects on atmospheric boundary layer stability, enhancement of air-sea gas exchange, and sea spray generation affecting heat and moisture fluxes. These interactions are particularly important for predicting storm development and intensity, as well as air-sea exchanges in climate models.

Coupling approaches include one-way forcing (traditional approach for operational forecasting), two-way coupling (wave models provide sea surface roughness and wave-induced stress to atmospheric models), and fully coupled Earth system models integrating atmospheric, wave, and ocean components \cite{Tolman2009,Janssen2004,Warner2010}. Applications include improving tropical cyclone intensity forecasts, enhancing storm surge predictions, incorporating wave effects in climate modeling, and improving weather forecasts for marine and coastal areas \cite{Chen2013,Bertin2015,Fan2012}.

\subsubsection{Wave-Sediment Coupling}

Waves play crucial roles in mobilizing, transporting, and depositing sediments in coastal environments. Coupling between waves and sediment processes is essential for understanding coastal morphodynamics \cite{Lesser2004}. Key processes include wave-induced bed shear stress and sediment mobilization, suspended sediment transport by wave orbital motion, longshore sediment transport driven by oblique wave approach, wave breaking and associated turbulence effects on sediment suspension, and feedback effects of changing bathymetry on wave transformation.

Coupling approaches include sequential coupling (wave models provide parameters to sediment transport models, which calculate morphological changes fed back to wave models) and process-based coupling (direct incorporation of sediment dynamics within wave models or unified frameworks). Morphological acceleration techniques bridge time scale gaps between hydrodynamic and morphological processes for long-term simulations \cite{Roelvink2009}. Applications include simulating beach evolution, assessing impacts of coastal structures on sediment transport and shoreline evolution, predicting storm-induced beach and dune erosion, and evaluating long-term coastal system response to sea level rise and changing wave climates \cite{Lesser2004,Roelvink2009}.

\subsubsection{Integrated Coastal Modeling Systems}

Comprehensive modeling frameworks combine multiple components including waves, currents, sediment transport, and water quality to address complex coastal problems \cite{Warner2010}. Major integrated systems include:

\textbf{Delft3D} integrates modules for waves (SWAN), hydrodynamics, sediment transport, morphology, and water quality, widely used for coastal and estuarine applications \cite{Lesser2004}.

\textbf{COAWST (Coupled Ocean-Atmosphere-Wave-Sediment Transport)} couples the ROMS ocean model, WRF atmospheric model, SWAN wave model, and Community Sediment Transport Model, designed for comprehensive coastal ocean simulations \cite{Warner2010}.

\textbf{MIKE} provides a suite of modeling tools including waves (MIKE 21 SW), hydrodynamics, sediment transport, and water quality with seamless integration capabilities, widely used in engineering applications \cite{DHI_site}.

\textbf{TELEMAC-MASCARET} includes modules for waves (TOMAWAC), hydrodynamics (TELEMAC-2D/3D), and sediment transport (SISYPHE), popular in European applications \cite{TELEMAC_site}.

\textbf{SCHISM+WWM} combines the unstructured-grid SCHISM hydrodynamic model with the WWM spectral wave model, designed for multi-scale coastal and estuarine applications with flexible mesh refinement \cite{Roland2012}.

While integrated modeling systems offer powerful capabilities, they present challenges including substantial computational demands for high-resolution or long-term simulations, ensuring consistency between model components in physical assumptions and numerical schemes, error propagation across coupled components, validation complexity requiring comprehensive multi-process datasets, and technical expertise requirements across multiple domains \cite{Warner2010,Bertin2015,Lesser2004}. Despite these challenges, integrated systems represent the state-of-the-art for comprehensive coastal and ocean modeling, with continued development improving efficiency, accuracy, and usability.

Despite significant advances in numerical wave modeling over the past five decades, fundamental challenges remain in physical process representation, computational efficiency, and predictive skill under extreme conditions. This section identifies major unsolved problems, critical research gaps, and promising directions for future development that will shape the next generation of wave modeling capabilities.

\subsection{Physical Process Representation Challenges}

\subsubsection{Wind Input and Whitecapping Dissipation}

The parameterization of wind input and whitecapping dissipation remains one of the most significant unsolved problems in wave modeling. These source terms govern wave growth and decay but are based on semi-empirical formulations with limited theoretical foundation \cite{Ardhuin2010}. Wind input parameterizations struggle to accurately represent momentum transfer from wind to waves across the full range of atmospheric stability conditions, wind speeds, and wave ages. The complexity of the turbulent atmospheric boundary layer and its interaction with the evolving wave field makes first-principles modeling computationally prohibitive for operational applications.

Whitecapping dissipation, which dominates wave energy decay in deep water, remains poorly understood despite decades of research. Current parameterizations rely heavily on empirical tuning parameters that lack clear physical interpretation and may not extrapolate reliably to conditions outside their calibration range \cite{Ardhuin2010}. The recent ST6 physics package represents progress through observation-based calibration, reducing errors by 15--25 percent in hurricane applications \cite{Liu2019}, but fundamental questions about the physics of wave breaking and energy dissipation remain unresolved. Future advances will require combining high-resolution field observations, laboratory experiments, and direct numerical simulation of breaking waves to develop more physically based parameterizations.

\subsubsection{Extreme Conditions and Strong Gradients}

Wave model performance degrades significantly under extreme conditions such as tropical cyclones, strong opposing currents, and rapidly varying wind fields. In tropical cyclones, the combination of extreme wind speeds, complex wind field structure, rapidly evolving conditions, and strong wave-current interaction challenges standard parameterizations developed for moderate conditions \cite{Tolman2009b}. Observations show that models often underpredict significant wave height in the most intense portions of hurricanes, with errors exceeding 2--3 meters in some cases.

Strong currents, particularly in regions like the Gulf Stream or Agulhas Current, can dramatically modify wave conditions through refraction, blocking, and enhanced steepness. When wave propagation opposes strong currents, waves can be blocked entirely or experience extreme steepening and breaking. Standard spectral models struggle to accurately represent these interactions, particularly when current gradients are sharp or when waves approach blocking conditions \cite{Ardhuin2017}. Phase-resolving models can capture some of these effects but remain computationally prohibitive for large-scale applications.

Steep or complex bathymetry in coastal regions violates the mild-slope assumptions and the homogeneous wave field underlying many wave transformation models. Submarine canyons, steep continental shelves, and complex reef structures create challenges for accurate wave prediction. The existing triad interaction models generally follow the weakly non-linear Hasselmann kinetic theory which assumes spatial homogeneity on O(100-km) scales that is not adequate for coastal modeling. While unstructured mesh capabilities provide geometric flexibility, fundamental limitations in the governing equations themselves restrict accuracy in regions where depth changes rapidly over distances comparable to the wavelength.

\subsubsection{Shallow Water Nonlinear Processes}

Nonlinear wave-wave interactions in shallow water, including triad interactions and higher-order effects, remain challenging to represent accurately and efficiently. The Lumped Triad Approximation (LTA) and Stochastic Parametric Model (SPM) used in operational models provide computational efficiency but sacrifice accuracy in some conditions \cite{Booij1999}. Observations in shallow water often show spectral evolution that differs from model predictions, particularly for bimodal seas or complex bathymetry. The existing triad interaction models generally follow the weakly non-linear Hasselmann kinetic theory, which assumes spatial homogeneity on O(100-km) scales that is not adequate for coastal modeling. Improved parameterizations or computationally efficient alternatives to current approximations are needed to improve the wave modeling in coastal areas.

Wave breaking in the surf zone involves complex turbulent processes that are parameterized rather than resolved in phase-averaged models. The widely used Battjes-Janssen model and its variants rely on empirical parameters that may not capture the full range of breaking behavior across different beach slopes, wave conditions, and bathymetric configurations \cite{Battjes1978}. Phase-resolving models can simulate breaking more directly but require empirical parameters for eddy viscosity or other turbulence closures. Fundamental understanding of surf zone turbulence and its interaction with wave propagation remains incomplete.

\subsection{Computational and Numerical Challenges}

The demand for higher resolution, larger domains, and longer simulations continues to push computational limits. Global wave forecasting at 1 km resolution, regional coastal modeling at 10 m resolution, and phase-resolving tsunami simulations at meter-scale resolution all require computational resources that exceed current operational capabilities. While high-performance computing continues to advance, wave model computational demands grow proportionally or faster as users seek finer resolution and more complex physics.

Multi-scale processes spanning orders of magnitude in space and time present fundamental challenges. Global swell propagation occurs over thousands of kilometers and days, while individual wave breaking occurs over meters and seconds. Efficiently coupling these scales within a single modeling framework remains difficult. Adaptive mesh refinement and nested grid approaches provide partial solutions but introduce complexity in boundary treatment, load balancing, and conservation properties \cite{Zijlema2010}.

Modern computing architectures emphasizing parallelism, accelerators, and heterogeneous systems require wave models to evolve beyond traditional CPU-based implementations. While GPU acceleration has demonstrated 20--50 times speedup for some models \cite{Shi2012,Yuan2024}, not all wave model components parallelize efficiently. Spectral integration, implicit solvers, and unstructured mesh operations present particular challenges for GPU implementation. Future model development must consider computational architecture from the outset rather than retrofitting existing codes.

\subsection{Data Assimilation and Uncertainty Quantification}

Current operational wave forecasting systems make limited use of available observations. While satellite altimeter data are routinely assimilated in some systems, the full potential of SAR imagery, HF radar, and in-situ spectral measurements remains underutilized. Advanced data assimilation techniques such as ensemble Kalman filters or variational methods have been explored but are not yet standard in operational wave forecasting \cite{Alves2013}. The high dimensionality of the wave model state space (spatial grid times spectral bins) and nonlinear dynamics pose challenges for efficient data assimilation.

Uncertainty quantification in wave predictions is increasingly important for decision-making but remains underdeveloped. Operational forecasts typically provide deterministic predictions without rigorous uncertainty estimates. Ensemble forecasting approaches can capture some uncertainty from atmospheric forcing but do not address model structural uncertainty or parameter uncertainty. Probabilistic wave forecasting systems that provide meaningful uncertainty information to end users require further development \cite{Alves2013}.

Multi-model ensembles combining predictions from different wave models offer potential to capture structural uncertainty but are computationally expensive and require careful interpretation. Different models may agree well under typical conditions but diverge significantly under extreme or unusual conditions where uncertainty is most critical. Systematic frameworks for combining multi-model predictions and quantifying their reliability across different conditions are needed.

\subsection{Emerging Research Frontiers}

Climate change impacts on wave conditions represent a growing application area requiring long-term simulations spanning decades to centuries. Efficient methods for multi-decadal wave climate projections under different emission scenarios are needed to support coastal adaptation planning. Waves may also provide feedbacks to the climate system through modifications of air-sea momentum, heat, and gas exchange, but these feedbacks are not yet well quantified or incorporated in Earth system models \cite{Hemer2013,Fan2012}.

The interaction between waves and coastal ecosystems is an emerging research area with implications for both ecosystem management and coastal protection. Coral reefs, seagrass beds, mangroves, and salt marshes modify wave conditions and provide natural coastal protection, but these effects are rarely included in operational wave models. Conversely, wave conditions affect ecosystem health, sediment dynamics, and habitat suitability. Developing coupled wave-ecosystem models that capture these bidirectional interactions will support integrated coastal management and nature-based solutions for coastal protection \cite{Lowe2009}.

Artificial intelligence (AI) and machine learning offer potential to improve wave modeling through data-driven parameterizations, efficient surrogate models, and enhanced data assimilation \cite{pokhrel2020forecasting}. Machine learning has been applied to predict wave conditions from atmospheric data, optimize source term parameters, and accelerate computationally expensive model components \cite{James2018}. However, fundamental questions remain about the reliability of data-driven approaches for extrapolation beyond training data, particularly for extreme events. Hybrid approaches combining physics-based models with machine learning components may offer the best path forward, leveraging the strengths of both approaches.

Integration of wave models into comprehensive multi-hazard early warning systems represents an operational frontier. Coastal communities face compound hazards from storm surge, waves, rainfall, and river flooding that interact in complex ways. Integrated systems that couple atmospheric, wave, surge, hydrologic, and inundation models can provide more complete hazard information than individual model components. Developing efficient coupling frameworks, managing computational demands, and communicating multi-hazard information to decision-makers and the public remain active areas of development \cite{Bertin2015}.

The next generation of wave models will likely feature tighter coupling with other Earth system components, more sophisticated physics parameterizations informed by machine learning and high-resolution process studies, efficient exploitation of emerging computing architectures, and comprehensive uncertainty quantification. Progress on these fronts will require continued collaboration between the wave modeling community, observational programs, computational scientists, and end users to ensure that model development addresses the most pressing scientific and societal needs.

\section{Conclusions}
Wave modeling has evolved from basic models to complex computational tools for forecasting, engineering, climate projection, and risk assessment worldwide. This review outlines the development, implementation, and application of modern wave models, aiding researchers in selecting suitable models and identifying areas for further research.

Wave modeling's progression from first to third generation models over five decades highlights advances in wave physics, computing, and observation technologies. Notable third generation models like SWAN, WAVEWATCH III, WAM, MIKE 21 SW, and TOMAWAC address energy transfer in waves while minimizing constraints. Recent innovations include the ST6 physics package, unstructured grids, and GPU use, reducing errors by 15-25\% and enabling high-resolution global forecasts. However, limitations in wind input, triad interactions, and whitecapping dissipation parameterization persist, relying on semi-empirical methods lacking a strong theoretical foundation.

Phase-resolving models like Boussinesq and non-hydrostatic types detail wave dynamics crucial for tsunami simulations, harbor resonance, and wave-structure interactions where phase information matters. In contrast, phase-averaged models statistically represent waves across scales, essential for forecasting, climate studies, and offshore engineering. The choice of wave models depends on resources and the nature of the problem. Hybrid models combining spectral models (for boundary conditions) and phase-resolving models (for detailed areas) are promising for efficient multi-scale wave modeling. Recent numerical techniques advances include finite-difference, finite-volume, and finite-element methods improving wave modeling accuracy and efficiency. Innovations like higher-order solutions, implicit integration, adaptive time stepping, and unstructured meshes (as in SWAN, WAVEWATCH III, TOMAWAC) significantly cut computational costs by about 5-10 times versus structured grids. Unstructured meshes enhance coastal geometry and bathymetry resolution, suggesting future models will leverage these advances for better accuracy and efficiency.

Advances in numerical techniques and GPU use have greatly impacted the field, significantly accelerating spectral models like WAM and WAVEWATCH III, and phase-resolving models like FUNWAVE-TVD, with 20-50x speedups compared to CPUs. Despite these gains, challenges persist in parallelizing certain wave model components for GPUs. Ongoing research is expected to optimize wave models for new platforms like TPUs and FPGAs.

Model validation using diverse data sources like buoys, satellite altimetry, SAR, and lab experiments confirms wave model reliability under typical open ocean conditions, with correlation coefficients over 0.9 and scatter indices under 0.2 for significant wave height. However, performance declines in coastal areas during extreme conditions, such as tropical cyclones and where strong current gradients exist, leading to underpredictions. Intercomparison projects by the WISE Group, JCOMM, and WMO highlight model strengths and weaknesses, guiding improvements. Wave models are crucial in coastal engineering, forecasting, climate studies, and marine energy. They provide design conditions, evaluate structures, assess erosion risks, and support marine safety and climate adaptation. They also enable tsunami modeling and wave energy assessments, demonstrating their importance in managing coastal hazards and ocean resources.

Cutting-edge coupled modeling frameworks integrate waves with currents, atmosphere, and sediment transport for coastal and ocean simulations, using systems like Delft3D and COAWST. These models are vital for morphodynamic studies, storm surge predictions, and coastal management, capturing interactions such as wave-induced currents and wave-driven sediment transport. Despite challenges in computation and validation, they offer precise multi-process predictions.

Current challenges include gaps in wind input, triad interactions and whitecapping dissipation understanding, performance issues under extreme conditions, high computational demands, and insufficient observation use and uncertainty quantification. Future research explores climate impacts on waves, wave-ecosystem interactions, AI for parameterizations, and multi-hazard systems. Next-gen wave models may feature tighter Earth system integration, physics-based parameterizations, GPU computing, and comprehensive probabilistic forecasts.

This review examines numerical wave modeling for coastal and oceanographic use, covering model formulations, numerical methods, validation, and practical applications. It identifies key models, compares their strengths and weaknesses, and offers guidance for model selection. This serves researchers, engineers, forecasters, and policymakers involved in wave modeling and management. Advances in physics, computing, and observation will boost our capacity to predict and manage wave processes amid rising coastal populations, marine activities, and climate change.

\section*{Acknowledgment}
This research was supported by the  U.S. Department. of the Navy, Naval Research Laboratory (NRL) under contract N00173-20-2-C007. The views expressed in this paper are solely those of the authors and do not necessarily reflect the views of the funding agency.

\section*{Declaration of generative AI and AI-assisted technologies in the manuscript preparation process}

During the preparation of this work, the author(s) used Writefull in order to improve English and polish it. After using this tool/service, the author(s) reviewed and edited the content as needed and take(s) full responsibility for the content of the published article.

\bibliographystyle{elsarticle-num}
\bibliography{new-ref,ref3-4}

\end{document}